\documentclass[11pt, peereview, onecolumn, letterpaper]{IEEEtran}

\usepackage{amsmath}
\usepackage{amssymb}
\usepackage{amsmath}
\usepackage{amsthm}
\usepackage{amssymb}
\usepackage{amscd}
\usepackage{latexsym}
\usepackage{multirow}
\usepackage{graphics}
\usepackage{color}
\usepackage{cite}
\usepackage{mathtools}
\usepackage{bbm}
\usepackage{amsfonts}
\usepackage[font=footnotesize]{caption}
\usepackage{setspace}

 \usepackage{caption}
\usepackage{subcaption}

\date{}
\newcommand{\eqdef}{\stackrel{\triangle}{=}}

\newtheoremstyle{mynewtheorem}
{5pt}
{5pt}
{\it}
{}
{\bf}
{.}
{.5em}
{\thmname{#1}\thmnumber{ #2}}%

\theoremstyle{mynewtheorem}
\newtheorem{myprop}{Proposition}

 \usepackage{graphicx}
 \usepackage{subcaption}
 \usepackage{mwe} 
 \DeclareCaptionLabelFormat{r-parens}{{#2}} 
\renewcommand{\thesubsection}{\Roman{section}.\Alph{subsection}}

\begin{document}

\title{\hfill{\small{\it Submitted to IEEE Transactions on Signal Processing}}\\
\vspace{1cm}\renewcommand{\baselinestretch}{1}\Huge Efficient L1-Norm Principal-Component Analysis via Bit Flipping}
 
\author{
		\thanks{A part of this paper was presented in  IEEE Intern. Conf. on Acoust., Speech, and  Signal Processing (ICASSP), Florence, Italy, May 2014 \cite{sandipan}. This work was supported in part by the National Science Foundation under Grant ECCS-1462341 and the Office of the Vice President for Research
			of Rochester Institute of Technology.}
Panos P. Markopoulos$^\dag$,
	Sandipan Kundu$^\triangledown$,
Shubham Chamadia$^\ddag$, and 
Dimitris A. Pados$^{\ddag *}$
\footnote{$^*$Corresponding author.}

\IEEEauthorblockA{
\vspace{0.2cm}
\vspace{-0.3 cm} $^\dag$Department  of Electrical and Microelectronic Engineering \\
\vspace{-0.3 cm} Rochester Institute of Technology\\
\vspace{-0.3  cm} Rochester, NY 14623 USA\\ 
\vspace{-0.3 cm} E-mail: \texttt{panos@rit.edu} \\
\vspace{0.2cm}
\vspace{-0.2 cm}  $^\triangledown$Qualcomm Technologies, Inc.\\
\vspace{-0.3  cm} San Jose, CA 95110 USA\\
\vspace{-0.3 cm} E-mail: \texttt{sandipan@qti.qualcomm.com} \\ 
\vspace{0.2cm}
 \vspace{-0.2 cm}  $^\ddag$Department of Electrical Engineering\\
\vspace{-0.3 cm} State University of New York at Buffalo\\
\vspace{-0.3  cm} Buffalo, NY 14260 USA\\
\vspace{-0.3 cm} E-mail: \texttt{\{shubhamc, pados\}@buffalo.edu} \\ 
 }
EDICS: MLR-ICAN, MLR-LEAR,  MLR-PATT, MDS-ALGO,  SSP-SSAN \\
 \vspace{-0.1cm}
Submitted: September 22, 2016
 	\vspace{-1.8cm}
}

\setlength{\textheight}{9in}
\pagestyle{plain}

\maketitle

\thispagestyle{empty}

\begin{abstract}
	 It was  shown recently that the $K$ L1-norm principal components (L1-PCs) of a real-valued data matrix $\mathbf X \in \mathbb R^{D \times N}$ ($N$ data samples of $D$ dimensions) can be exactly calculated   with  cost  $\mathcal{O}(2^{NK})$ or, when advantageous,   $\mathcal{O}(N^{dK - K + 1})$ where $d=\mathrm{rank}(\mathbf X)$, $K<d$ \cite{Panos1,Panos2}. 
	In  applications where   $\mathbf X$ is large (e.g., ``big" data of large $N$ and/or ``heavy" data of large $d$),  these costs are prohibitive.
	In this work, we
 present a novel  suboptimal algorithm for the calculation of the $K < d$ L1-PCs of $\mathbf X$ of cost $\mathcal O (ND \mathrm{min} \{ N,D\}  + N^2(K^4 + dK^2) + dNK^3)$, which is comparable  to that of standard (L2-norm) PC analysis.
Our theoretical  and  experimental studies show that  the proposed algorithm calculates the exact optimal L1-PCs  with high  frequency and achieves  higher value in the L1-PC optimization metric than any known alternative algorithm of comparable computational cost.
The superiority of the  calculated L1-PCs over standard L2-PCs (singular vectors) in characterizing   potentially faulty data/measurements is demonstrated with experiments on data dimensionality reduction
and disease diagnosis from  genomic data.
\end{abstract}

{\bf \emph{Index Terms}} --- Dimensionality reduction, data analytics, eigen-decomposition, L1 norm, L2 norm, machine learning, outlier resistance, principal component analysis, subspace signal processing. \vspace{-0.3cm}

\section{Introduction }

\label{ch2_sec1}
Principal-Component Analysis  (PCA)    \cite{Pearson}   has been  a ``mainstay" of   signal processing, machine learning, pattern recognition, and classification \cite{Jolliffe, Bishop, Duda} for more than a century. 
Numerous important applications of PCA can be found in the fields of  wireless communications, computer networks,  computer
vision,  image processing,  bio-informatics/genomics, and neuroscience, to name a few. 
Broadly speaking, PCA seeks to calculate  orthogonal   directions  that define a  subspace wherein data presence is maximized. 
Traditionally,    PCA quantifies data presence by the Frobenius (L2) norm of the  projected data onto the subspace, or, equivalently, the Euclidean distance of the original data from their subspace representations (L2-norm of residual error). 
Herein, we will be referring to  standard PCA as L2-PCA. 
Key strengths of L2-PCA  are (i) its low-complexity implementation (quadratic in the number of data points)  by means of  singular-value decomposition (SVD) of the data matrix and (ii) the reliable approximation that it offers to the nominal principal  data subspace, when calculated over sufficiently many  clean/nominal  or benign-noise corrupted  data points.

In the advent of the big-data era, datasets   often include grossly corrupted, highly deviating, irregular data points (outliers),
due to a variety of   causes such as  transient sensor malfunctions, errors in data
transmission/transcription, errors in training data labeling, and  bursty-noise data corruption, to name a few  \cite{Barnett, Tukey}.
Regretfully, standard L2-PCA is well-known to be fragile in the presence of such faulty data, even when they appear in a vanishingly small fraction  of the training set   \cite{Candes}. %
The reason is that the L2-norm objective of standard PCA (minimization of error variance or maximization of squared projection magnitude) gives  squared importance on the magnitude of  every datum, thus overemphasizing peripheral data points.

To remedy the impact of outliers, researchers from the fields of data analysis and  signal processing have long focused on calculating   subspaces of minimum absolute error deviations, instead of minimum error variances.
Important early theoretical studies date   back  to the 1940s \cite{Singleton, Karst, Barrodale1, Barrodale2}.
In the past decade, there have been several L1-norm-based principal component calculators under the general label ``L1-PCA"  \cite{Ke2003,Ke2005,Brooks2013a,Brooks2013b, Eriksson2010,He2011,Yu2012,Nie2011,Kwak2008,Kwak2009,Li2009, PanosL1fusion, Pang2010,Ding2006, McCoy2011, Funatsu2010,Meng2012, Wang2012a, Wang2012b, NIKOSDOA, PanosDOASPIE, Johnson2014, Markopoulos2016, sandipan}. 
In these  works, researchers seek  either  to (i)  minimize the aggregate absolute data representation error or  (ii)  maximize the aggregate absolute magnitude of the projected data points. 
For approach (i), it has been shown that the error surface is non-smooth and the
problem non-convex, resisting attempts to reach an exact solution
even with exponential computational cost \cite{Nesterov2013}. Therefore, only suboptimal algorithms exist in the literature \cite{Ke2003, Ke2005, Brooks2013a, Brooks2013b}. 
For approach (ii), a problem known as  \emph{ maximum-projection L1-PCA}, the works in \cite{Panos1, Panos2} proved  that maximum-projection L1-PCA is not   NP-hard for fixed data dimension $D$  and offered the  first two optimal algorithms in the literature for   exact calculation.  
Specifically, for a data record of size $N$, $\mathbf X \in \mathbb R^{D \times N}$, the  first optimal algorithm solves L1-PCA exactly with complexity  $\mathcal{O}(2^{NK})$; the second optimal algorithm solves L1-PCA exactly with complexity $\mathcal{O}(N^{\mathrm{rank}(\mathbf X)K - K + 1})$ where $K < \mathrm{rank}(\mathbf X)$ is the desired number of L1-PCs. 
Before the results of \cite{Panos1, Panos2},  several  low-complexity  suboptimal algorithms for  L1-PCA existed in the literature \cite{Kwak2008, Kwak2009, Nie2011,McCoy2011}.  However, in lack of the optimal solution, no absolute performance evaluation  of those algorithms with respect to the L1-PCA metric could be conducted until \cite{Panos1, Panos2} made  the optimal value available. 
Today, optimal-solution-informed experimental studies  indicate that the existing low-cost approximate algorithms for L1-PCA  yield non-negligible performance degradation, in particular when more than one principal component is calculated.

In this present work, we
introduce  \emph{L1-BF}, a bit-flipping based algorithm for the calculation of the $K$ L1-PCs of any rank-$d$ data matrix $\mathbf X \in \mathbb R^{D \times N}$  with complexity $\mathcal O (ND \mathrm{min} \{ N,D\} + N^2(K^4 + dK^2) + NdK^3 )$.
The proposed  algorithm is accompanied by:  (i) Formal  proof of convergence and theoretical analysis of   converging points; (ii) detailed asymptotic  complexity derivation;  (iii) theoretically proven   performance guarantees; (iv) L1-PCA performance comparisons with 
state-of-the-art counterparts of comparable computational cost \cite{Kwak2008, Kwak2009, Nie2011,McCoy2011};  and (v) outlier-resistance experiments on dimensionality reduction,  foreground motion detection in surveillance videos, and disease diagnosis from genomic data. Our studies show that L1-BF outperforms all suboptimal counterparts of comparable cost with respect to the L1-PCA metric and retains high outlier-resistance similar to that of optimal L1-PCA. Thus, the proposed algorithm may bridge the gap between computationally efficient and outlier-resistant  principal component analysis. 

The rest of the paper is organized as follows.  
Section II presents the problem statement and reviews briefly  pertinent technical background. 
Section III is devoted to the development and analysis of the proposed algorithm. 
Extensive experimental studies  are presented in Section IV. A
few concluding remarks are drawn in Section V.

\section{Problem Statement and Background}

\label{ch2_sec2}

Given a data matrix ${\mathbf X} = [\mathbf x_{1}, \mathbf x_{2}, \ldots, \mathbf x_{N}]\in \mathbb R^{D \times N}$ of rank $d \leq \min \{D,N\}$, we are interested in calculating a low-rank data subspace of dimensionality $K < d$ in the form of an orthonormal basis $\mathbf Q_{L1}  \in \mathbb R^{D \times K}$ that solves
\begin{align}
{
\mathbf Q_{L1} = \underset{ 
\begin{smallmatrix}
\mathbf Q = [\mathbf q_{1}, \ldots, \mathbf q_{K}] \in \mathbb R^{D \times K}  \\
\mathbf Q^\top \mathbf Q = \mathbf I_{K}
\end{smallmatrix}
}{\mathrm{argmax}}~ \sum_{k=1}^{K}\left\| \mathbf X^\top \mathbf q_{k} \right\|_1. 
\label{ch2_l1pca}
}
\end{align}
 In \eqref{ch2_l1pca}, $\left\|\cdot\right\|_1$ denotes the element-wise L1-norm of the vector/matrix argument that  returns the sum of the absolute values of the individual entries.
\cite{Panos1, Panos2} and \cite{McCoy2011} presented different proofs  that L1-PCA in the form of \eqref{ch2_l1pca} is  formally NP-hard problem in \emph{jointly asymptotic} $N$, $d$. 
Suboptimal algorithms (with  non-negligible performance degradation) for approximating the solution in   \eqref{ch2_l1pca} were proposed in \cite{Kwak2008, Kwak2009, Nie2011, McCoy2011}.
 \cite{Panos1, Panos2} showed for the first time that for fixed data dimension $D$,  \eqref{ch2_l1pca} is not NP-hard and presented two optimal algorithms for solving it.

Below,  we review  briefly the optimal solution to \eqref{ch2_l1pca} and the suboptimal algorithms that exist in the literature.

\subsection{Optimal Solution}
For any matrix $\mathbf A \in \mathbb R^{m \times n}$,   $m>n$, that admits SVD $\mathbf A = \mathbf U \Sigma_{n \times n} \mathbf V^T$,   define $U(\mathbf A) \eqdef \mathbf U \mathbf V^\top$.  
Let $\|  \cdot \|_*$ denote nuclear norm. 
In   \cite{Panos1, Panos2} it was shown that, if 
\begin{equation}
 {\bf B}_\mathrm{{opt}} =  \underset{{\bf B}\in\{\pm1\}^{N\times K}}{\mathrm{argmax}}\left\|{\bf X}{\bf B}\right\|_*,
\label{ch2_nucnorm}
\end{equation}
then
\begin{align}
{\mathbf Q}_{L1} =   U({\bf X}{\bf B}_\mathrm{{opt}})
\label{ch2_btoq}
\end{align}
 is a solution   to \eqref{ch2_l1pca}. In addition, $\left\|{\mathbf Q}_{L1}^\top{\bf X}\right\|_1=\left\|{\bf X}{\bf B}_\mathrm{{opt}}\right\|_*$ and $\mathbf B_{\mathrm{opt}} = \mathrm{sgn}(\mathbf X^\top \mathbf Q_{L1})$ \cite{Panos1, Panos2}.

For $K=1$,    \eqref{ch2_nucnorm} takes the  binary quadratic form\footnote{For every $\mathbf a \in \mathbb R^{d}$, the nuclear and euclidean norm trivially coincide, i.e.,  $\left\| \mathbf a\right\|_* =\left\| \mathbf a\right\|_2$.}  
\begin{align}
\mathbf b_{\mathrm{opt}} = \underset{\mathbf b \in \{\pm 1\}^{N}}{\mathrm{argmax}}~~\left\| \mathbf X \mathbf b\right\|_2^2
\label{ch2_quad}
\end{align}
and the L1-principal component of $\mathbf X$ is given  by
\begin{align}
\mathbf q_{L1} = U(\mathbf X \mathbf b_{\mathrm{opt}}) =\mathbf X \mathbf b_{\mathrm{opt}}\left\| \mathbf X \mathbf b_{\mathrm{opt}}\right\|_2^{-1} . 
\label{btoq}
\end{align}
In addition, $\left\| \mathbf X^\top \mathbf q_{L1}\right\|_1 =\left\| \mathbf X  \mathbf b_{\mathrm{opt}}\right\|_* = \left\| \mathbf X  \mathbf b_{\mathrm{opt}}\right\|_2$ and $\mathbf b_{\mathrm{opt}} = \mathrm{sgn} (\mathbf X^\top \mathbf q_{L1})$.

In view of \eqref{ch2_nucnorm}, \eqref{ch2_btoq}, the first optimal algorithm in \cite{Panos1,Panos2} performs an exhaustive search over the  size-$2^{NK}$ candidate set $\{ \pm 1 \}^{N \times K}$ to obtain a solution $\mathbf B_{\mathrm{opt}}$   to \eqref{ch2_nucnorm}; then $\mathbf Q_{L1}$ is returned by SVD on $\mathbf X \mathbf B_{\mathrm{opt}}$.\footnote{In practice, the exhaustive-search optimal algorithm takes advantage of the nuclear-norm invariability to negations and permutations of the columns of the matrix argument and searches  exhaustively in a size-$\binom{2^{N-1}+K-1}{K}$ subset of $\{ \pm  1\}^{N \times K}$ wherein a solution to \eqref{ch2_nucnorm} is guaranteed to exist.} The second optimal algorithm in \cite{Panos1,Panos2}, of polynomial complexity,  constructs and searches inside a subset of $\{ \pm 1\}^{N \times K}$ wherein a solution to \eqref{ch2_nucnorm} is proven to exist. Importantly, for $d$ constant  with respect to $N$, the cost to construct and search exhaustively within this  set is $\mathcal O(N^{dK-K+1})$.

From  \eqref{ch2_nucnorm}-\eqref{btoq}, it is seen that,   in contrast to L2-PCA, in L1-PCA the scalability principle does not hold. 
That is,  $\mathbf q_{L1} = \underset{\mathbf q \in \mathbb R^{D \times 1}; ~\| \mathbf q\|_2=1}{\mathrm{argmax}}~\| \mathbf X^\top \mathbf q\|_1
$
is not (in general) a column of $\mathbf Q_{L1} = 
\underset{\mathbf Q \in \mathbb R^{D \times K}; ~ \mathbf Q^\top \mathbf Q = \mathbf I_{K}}{\mathrm{argmax}}~\| \mathbf X^\top \mathbf Q\|_1 
$
for some $K>1$.
Therefore, the size-$(K>1)$ L1-PCA problem cannot be  translated into a series of size-$(K=1)$ L1-PC problems simply by projecting the data-matrix onto the null-space of the previous solutions. 

\vspace{-0.1cm}
\subsection{State-of-the-art Approximate Algorithms}

\label{ch2_sec4}

\subsubsection{Fixed-point Iterations with Successive Nullspace Projections  \cite{Kwak2008, Kwak2009}}

Kwak et al. \cite{Kwak2008, Kwak2009} made an important early contribution to the field by proposing a fixed point (FP) iteration  to approximate  the $K=1$ L1-PC solution. Following our formulation and notation in Section \ref{ch2_sec2}, the  algorithm has the iterative form 
\begin{align}
\mathbf b^{(t)} = \mathrm{sgn} \left( \mathbf X^\top \mathbf X \mathbf b^{(t-1)} \right), ~t=2,3,\ldots,
\label{ch2_fpi}
\end{align}
where $\mathbf b^{(1)}$ is an arbitrary initialization point in $\{ \pm 1\}^{N}$. For any initialization, \eqref{ch2_fpi} is guaranteed to converge to a fixed point of $\mathrm{sgn} \left( \mathbf X^\top \mathbf X \mathbf b \right)$ in $\Phi (\mathbf  X)  = \{ \mathbf b \in \{\pm 1\}^N:~\mathbf b = \mathrm{sgn}(\mathbf X^\top \mathbf X \mathbf b)\}$.
Then,  the L1-PC  can be approximated by 
$
\mathbf q_{\mathrm{fp}} =  \mathbf X   \mathbf b_{\mathrm{fp}} \| \mathbf X \mathbf b_{\mathrm{opt}}\|_2^{-1}
$
 where $\mathbf b_{\mathrm{fp}}$ is the converging point of the iteratively generated   sequence  $\{ \mathbf b^{(t)} \}$.
For $K>1$,   L1-PCs are approximated sequentially  in a greedy fashion. 
That is, the $k$th L1-PC $\mathbf q_{\mathrm{fp}, k}$, $k>1$, is calculated  by the above procedure, having 
replaced  $\mathbf X$ by its projection onto the nullspace of the previously calculated components, $(\mathbf I_{D} - \sum_{i=1}^{k-1} \mathbf q_{\mathrm{fp}, i}\mathbf q_{\mathrm{fp}, i}^\top) \mathbf X$. 
The complexity of this algorithm is $\mathcal O(MNDK)$ where $M$ is the maximum number of iterations per component.
Considering $M$ to be bounded by a linear function of $N$, or practically terminating the iterations at most at  $N$, the complexity of this algorithm can be expressed as $\mathcal O(N^2DK)$.

\subsubsection{Iterative Alternating Optimization (Non-greedy) \cite{Nie2011}}
Nie et al. \cite{Nie2011} offered a significant advancement for the case $K>1$. Following, again, our formulation and notation, they  proposed a converging iterative algorithm that, initialized at an arbitrary orthonormal matrix  $\mathbf Q^{(1)} \in \mathbb R^{D \times K}$, calculates 
\begin{align}
\mathbf B^{(t)} = \mathrm{sgn} \left( \mathbf X^\top \mathbf Q^{(t-1)}\right)  \text{ ~~and~~ }
\mathbf Q^{(t)} = U \left(  \mathbf X \mathbf B^{(t)} \right),~t=2, 3, \ldots.
	\label{arlington}
\end{align}
The solution to \eqref{ch2_l1pca} is approximated by the convergence   point of the   generated sequence  $\{ \mathbf Q^{(t)} \}$, say $ {\mathbf Q}_{\mathrm{ao}}$. Notice that, for $K=1$, the algorithm coincides with the one in \cite{Kwak2008}.
The computational complexity of \eqref{arlington} is  $\mathcal O(T(ND + K^2))$ where $T$ is the maximum number of iterations per component.  Considering $T$ to be bounded by a linear function of $NK$, or practically terminating the iterations at most at $NK$, the complexity of the algorithm becomes $\mathcal O(N^2DK + NK^3)$.

\subsubsection{SDP with Successive Nullspace Projections \cite{McCoy2011}}
McCoy and Tropp \cite{McCoy2011} suggested a novel semi-definite programming (SDP) view of the problem.  The binary-optimization problem in \eqref{ch2_quad} can be rewritten as
\begin{align}
\underset{
	\begin{smallmatrix}
\mathbf Z \in \mathcal S_{+}^{N},~[\mathbf Z]_{n,n}=1 ~\forall n \\
\mathrm{rank}(\mathbf Z)=1
	\end{smallmatrix} }{\mathrm{maximize}}~\mathrm{Tr}\left( \mathbf Z \mathbf X^\top \mathbf X\right) 
\label{ch2_sd}
\end{align}
where $\mathcal S_{+}^{N}$   the set of positive semi-definite matrices in $\mathbb R^{N \times N}$. Specifically, if   $\mathbf Z_{\mathrm{opt}}$ is the solution to  \eqref{ch2_sd}, then any column of $\mathbf Z_{\mathrm{opt}}$ is a solution to  \eqref{ch2_quad}. Then,  the algorithm   relaxes the non-convex rank constraint in \eqref{ch2_sd}  and finds instead the solution $\mathbf Z_{\mathrm{sdp}}$ to the convex semi-definite program 
\begin{align}
\hspace{-1cm}
\underset{\mathbf Z \in \mathcal S_{+}^{N},~ [\mathbf Z]_{n,n}=1~ \forall n}{\mathrm{maximize}}~\mathrm{Tr}\left( \mathbf Z \mathbf X^\top \mathbf X\right).
 \label{ch2_sdp}
\end{align}
In this present paper's notation, to obtain an approximation to $\mathbf b_{\mathrm{opt}}$, say $ {\mathbf b}_{\mathrm{sdp}}$,  the algorithm factorizes ${\mathbf Z}_{\mathrm{sdp}} = {\mathbf W} \mathbf W^\top$,  $\mathbf W \in \mathbb R^{N \times N}$, and calculates $L$ instances of ${\mathbf b}= \mathrm{sgn}\left(\mathbf W^\top \mathbf a \right)$ for vectors $\mathbf a$ drawn from $\mathcal N (\mathbf 0_{N}, \mathbf I_{N})$ ($L$-instance Gaussian randomization). $ {\mathbf b}_{\mathrm{sdp}}$ is chosen to be the instance that maximizes $\left\| \mathbf X \mathbf b\right\|_2$ and the solution to \eqref{ch2_l1pca} is approximated by $ {\mathbf q}_{\mathrm{sdp}}  =\mathbf X  {\mathbf b}_{\mathrm{sdp}} \left\| \mathbf X   {\mathbf b}_{\mathrm{sdp}} \right\|_2^{-1}$.
For $K>1$, similar to \cite{Kwak2008}, \cite{McCoy2011} follows the method of sequential nullspace projections  calculating the $k$th L1-PC $ {\mathbf q}_{\mathrm{sdp},k}$  as the L1-PC of $(\mathbf I_{D} - \sum_{i=1}^{k-1}  {\mathbf q}_{\mathrm{sdp},i} {\mathbf q}_{\mathrm{sdp},i}^\top) \mathbf X$.
The complexity to solve within $\epsilon$ accuracy the SDP in \eqref{ch2_sdp} is $\mathcal O( N^{3.5} \mathrm{log}(1/\epsilon) )$ \cite{Luo2010}. Thus, the overall computational cost  of the algorithm  is $\mathcal O(KN^{3.5} \mathrm{log}(1/\epsilon) + KL(N^2 + DN))$.

\section{Proposed Algorithm }%

\label{ch2_sec5}

We begin our new algorithmic developments with the  following Proposition.

\begin{myprop}
Assume that   data matrix $\mathbf X$ admits compact singular-value decomposition   $\mathbf X \overset{\mathrm{SVD}}{=} \mathbf U_{D \times d} \mathbf \Sigma_{d\times d} \mathbf V_{N \times d}^\top \in \mathbb R^{D \times N}$ where  $d=\mathrm{rank}(\mathbf X) \leq \min \{D,N \}$ and define 
\begin{align}
\mathbf Y = [\mathbf y_{1}, \mathbf y_{2}, \ldots, \mathbf y_{N}] \eqdef  {\mathbf \Sigma} \mathbf V^\top \in \mathbb R^{d \times N}.
\label{ch2_defY}
\end{align}
Then, for any $\mathbf B \in \{ \pm 1\}^{N \times K}$,  $\left\| \mathbf X \mathbf B\right\|_* =\left\| \mathbf Y \mathbf B\right\|_*$. 
 \label{ch2_fromXtoY}
  \end{myprop}
\noindent \emph{Proof:}
Notice that
\begin{align}
\mathbf X^\top \mathbf X & = \mathbf V \mathbf \Sigma \mathbf U^\top \mathbf U \mathbf \Sigma \mathbf V^\top.
\label{ch2_XtoY}
\end{align}
Thus, $\left\| \mathbf X \mathbf B\right\|_* = \mathrm{Tr}\left( \sqrt{\mathbf B^\top \mathbf X^\top \mathbf X   \mathbf B} \right)  = \mathrm{Tr}\left( \sqrt{\mathbf B^\top \mathbf Y^\top \mathbf Y  \mathbf B} \right) =\left\| \mathbf Y \mathbf B\right\|_* $ where for any positive semi-definite  symmetric matrix  $\mathbf A \in \mathbb R^{m \times m}$, $\sqrt{\mathbf A} \sqrt{\mathbf A} = \mathbf A$.
\hfill $\square$
 
By  Proposition \ref{ch2_fromXtoY}, the proposed algorithm will attempt to find a solution to  \eqref{ch2_nucnorm}  by solving instead the reduced-size problem 
\begin{align}
\mathbf B_{\mathrm{opt}} = \underset{\mathbf B \in \{ \pm 1\}^{N \times K}}{\mathrm{argmax}}\left\| \mathbf Y \mathbf B \right\|_*,
\label{ch2_nucnormmet}
\end{align}
which  for $K=1$  takes the form 
\begin{align}
\mathbf b_{\mathrm{opt}} = \underset{\mathbf b \in \{ \pm 1\}^{N \times 1}}{\mathrm{argmax}}\left\| \mathbf Y \mathbf b\right\|_2^2. 
\label{ch2_quad2}
\end{align}
This problem-size reduction, with no loss of optimality  so far, contributes to the low cost of the proposed algorithm. In the sequel, we present separately the cases  $K=1$ and  $K>1$.

\subsection{Calculation of the L1-Principal Component (K = 1)}
\label{single_component}
 
 First, we attempt to find efficiently a quality approximation to $\mathbf b_{\mathrm{opt}}$ in  \eqref{ch2_quad2}, say  $\mathbf b_{\mathrm{bf}}$, by a bit-flipping search procedure. 
Then, per \eqref{btoq}, 
\begin{align}
\mathbf q_{\mathrm{bf}} =    \mathbf X \mathbf b_{\mathrm{bf}}\left\| \mathbf X \mathbf b_{\mathrm{bf}}\right\|_2^{-1}
\label{ch2_bbftoqbf}
\end{align}
will be the suggested L1-principal component of the data. 
Proposition \ref{ch2_degbound}, presented herein for the first time, bounds the L1 error metric of interest by L2-norm differences. 
\begin{myprop} \label{ch2_degbound}
For any $ {\mathbf b} \in \{ \pm 1\}^{N}$ and corresponding  $ {\mathbf q} = \mathbf X  {\mathbf b}\left\| \mathbf X  {\mathbf b}\right\|_2^{-1}$, 
\begin{align}
\left\| \mathbf X^\top \mathbf q_{L1}\right\|_1 -\left\|  \mathbf X^\top  {\mathbf q}\right\|_1 \leq\left\| \mathbf Y \mathbf b_{\mathrm{opt}}\right\|_2 -\left\| \mathbf Y  {\mathbf b} \right\|_2 
\end{align}
 with equality if $\mathbf b = \mathrm{sgn} (\mathbf Y^\top \mathbf Y \mathbf b)  {=} \mathrm{sgn} (\mathbf X^\top \mathbf X \mathbf b).$  
 \end{myprop} 
\noindent \emph{Proof:}
By \eqref{ch2_nucnorm}-\eqref{btoq} and Proposition \ref{ch2_fromXtoY}, 
\begin{align}
& \left\| \mathbf X^\top \mathbf q_{L1}\right\|_1 -\left\|  \mathbf X^\top  {\mathbf q}\right\|_1    =  \left\| \mathbf X \mathbf b_{\mathrm{opt}}\right\|_2 - \left\|  \mathbf X^\top \mathbf X  {\mathbf b} \right\|_1\left\| \mathbf X  {\mathbf b}\right\|_2^{-1} \nonumber  \\
& =  \left\| \mathbf Y \mathbf b_{\mathrm{opt}}\right\|_2 - \left\|  \mathbf Y^\top \mathbf Y  {\mathbf b} \right\|_1\left\| \mathbf Y  {\mathbf b}\right\|_2^{-1}  \nonumber  \\
&=  \left\| \mathbf Y \mathbf b_{\mathrm{opt}}\right\|_2 -  (\mathrm{max}_{\mathbf z \in \{ \pm 1\}^{N} } \mathbf z^\top \mathbf Y^\top \mathbf Y  {\mathbf b})\left\| \mathbf Y  {\mathbf b} \right\|_2^{-1} \nonumber  \\
& \leq \left\| \mathbf Y \mathbf b_{\mathrm{opt}}\right\|_2 -     {\mathbf b}^\top \mathbf Y^\top \mathbf Y  {\mathbf b}\left\| \mathbf Y  {\mathbf b} \right\|_2^{-1} \nonumber  \\
&= \left\| \mathbf Y \mathbf b_{\mathrm{opt}}\right\|_2 -    \left\| \mathbf Y  {\mathbf b} \right\|_2.  
\end{align}
Equality holds when $\mathbf b = \mathrm{sgn} (\mathbf Y^\top \mathbf Y \mathbf b) =  {\mathrm{argmax}}_{\mathbf z \in \{ \pm 1\}^N } \mathbf z^\top \mathbf Y^\top \mathbf Y \mathbf b$.
\hfill $\square$

In particular, by Proposition \ref{ch2_degbound},  the performance degradation in the L1-PCA metric $\| \mathbf X^\top \mathbf q\|_1$  when  $  {\mathbf q}  = \mathbf X  {\mathbf b}\left\| \mathbf X  {\mathbf b}\right\|_{2}^{-1}$ is used instead of the optimal $\mathbf q_{L1}$ is upper bounded by the performance degradation  in the metric of \eqref{ch2_quad2} when $ {\mathbf b}$ is used instead of the optimal $\mathbf b_{\mathrm{opt}}$. 

In view of Proposition \ref{ch2_degbound}, the proposed algorithm operates as follows. 
The algorithm initializes at a binary vector ${\mathbf b^{(1)}}$ and employs bit-flipping (BF) iterations  to  generate  a sequence of binary vectors $\{ \mathbf b^{(t)}\}$.
At each iteration step, the algorithm browses all bits that have not been flipped before kept in an index-set\footnote{The index-set  $\mathcal L$ is used to restrain the ``greediness" of unconstrained single-bit flipping iterations.} $\mathcal L$. Then, the algorithm  negates (flips) the single bit in $\mathcal L$ that, when flipped,  offers the highest increase in the metric of \eqref{ch2_quad2}.
If  no   bit exists in $\mathcal L$  the negation of which  increases the quadratic metric \eqref{ch2_quad2}, then  $\mathcal L$ is reset to  $\{1, 2, \ldots, N \}$  and all $N$ bits become eligible for flipping consideration again. 
The iterations terminate  when   metric \eqref{ch2_quad2} cannot be further increased by    any single-bit flipping.

   Mathematically,  at  $t$th iteration step,  $t \geq 1$,
  the   optimization objective  \eqref{ch2_quad2} is
\begin{equation}
\left\| \mathbf Y  \mathbf b^{(t)}\right\|_{2}^2  = \| \mathbf Y\|_F^2  + \hspace{-0.2cm} \sum_{
\begin{smallmatrix} n \in \{1, 2, \ldots, N \} \\
m \in \{1, 2, \ldots, N \} \setminus n
\end{smallmatrix}}    b_n^{(t)}  b_m^{(t)} \mathbf y_n^\top \mathbf y_m
\label{ch2_quadval}  
\end{equation} 
where $\| \mathbf Y \|_F $ is the Frobenius (Euclidean) norm of $\mathbf Y$.
Factorizing    \eqref{ch2_quadval}, we observe that 
the \emph{contribution} of the $n$th bit of $\mathbf b^{(t)}$, $b_n^{(t)}$,  to \eqref{ch2_quadval} is 
\begin{align}
\alpha\left( \mathbf b^{(t)}, n \right) &=   2 b_{n}^{(t)}  \sum_{ m \in \{1, 2, \ldots, N \} \setminus n}   \,  b_m^{(t)} \mathbf y_n^\top \mathbf y_m
\nonumber  = 2 \left( b_{n}^{(t)} \mathbf y_{n}^\top \mathbf Y \mathbf b^{(t)} - \| \mathbf y_{n}\|_2^2 \right). 
\label{ch2_contribution} 
\end{align}
That is, for any $n \in \{1, 2, \ldots, N \}$,  \eqref{ch2_quadval} can be written as 
\begin{align}
\left\| \mathbf Y  \mathbf b^{(t)}\right\|_{2}^2  = \alpha\left( \mathbf b^{(t)}, n \right) +  \beta \left( \mathbf b^{(t)},n \right)
\end{align}
 where  $\beta \left( \mathbf b^{(t)},n \right)$ is a constant with respect to $b_{n}^{(t)}$.
Then,  if we flip the $n$th bit by setting $\mathbf b^{(t+1)} = \mathbf b^{(t)} - 2 b_{n}^{(t)}\mathbf e_{n,N}$ (where $\mathbf e_{n,N}$ is the $n$th column of the size-$N$ identity matrix $\mathbf I_{N}$),  the change to the quadratic metric  in \eqref{ch2_quad2} is
\begin{align}
 \left\| \mathbf Y (\mathbf b^{(t)} - 2 b_{n}^{(t)}\mathbf e_{n,N})\right\|_2^2 -\left\| \mathbf Y \mathbf b^{(t)} \right\|_2^2 = -2 \alpha\left( \mathbf b^{(t)}, n \right). 
 \label{ch2_metricincrease}
\end{align} 
Thus,   if the contribution of the $n$th bit to the quadratic of \eqref{ch2_quadval}, $\alpha\left( \mathbf b^{(t)}, n \right)$, is negative, flipping  $b_n^{(t)}$ will increase the quadratic by $2|\alpha\left( \mathbf b^{(t)}, n \right)|$, whereas if $\alpha\left( \mathbf b^{(t)}, n \right) $ is positive, flipping $b_n$ will decrease the quadratic by $2|\alpha\left( \mathbf b^{(t)}, n \right)|$.  

In view of this analysis,  L1-BF  is initialized at an  arbitrary binary antipodal vector $\mathbf b^{(1)}$ and $\mathcal L = \{1, 2, \ldots, N \}$. The initial bit contributions $\alpha \left(\mathbf b^{(1)},1 \right), \ldots, \alpha \left(\mathbf b^{(1)}, N \right)$ are calculated by \eqref{ch2_contribution}. Then, at iteration step $t \geq 1$, we find
\begin{equation}
n = \underset{m \in \mathcal L}{\mathrm{argmin}}~~\alpha\left( \mathbf b^{(t)}, m \right).
 \label{ch2_update2}
 \end{equation}
 If     $\alpha\left( \mathbf b^{(t)}, n \right)<0$,  $b_n^{(t)}$ is flipped by  setting  
$ \mathbf b^{(t+1)} = \mathbf b^{(t)} - 2 b_{n}^{(t)} \mathbf e_{n,N}$,   $\alpha \left(\mathbf b^{(t+1)},1 \right)$, $\ldots$, $\alpha \left(\mathbf b^{(t+1)},N \right)$ are recalculated,  
  and   $\mathcal L$ is updated to $\mathcal L \setminus n$.
If, otherwise,  $\alpha\left( \mathbf b^{(t)}, n \right) \geq 0$, a new solution $n$ to
\eqref{ch2_update2} is obtained after resetting $\mathcal L$  to $\{1, 2, \ldots, N \}$. 
If, for the new solution $n$,    $\alpha\left( \mathbf b^{(t)}, n \right) < 0$, then $b_n^{(t)}$ is flipped,   $\alpha \left(\mathbf b^{(t+1)},1 \right), \ldots, \alpha \left(\mathbf b^{(t+1)},N \right)$ are recalculated,  and     $\mathcal L $ is updated to $  \mathcal L \setminus  n $.
 Otherwise, the iterations terminate and the algorithm returns $\mathbf b_{\mathrm{bf}} = \mathbf b^{(t)}$,  $\mathbf q_{\mathrm{bf}}  = \mathbf Y \mathbf b_{\mathrm{bf}}\left\| \mathbf Y \mathbf b_{\mathrm{bf}} \right\|_2^{-1}$. 
 
Notice that   the contribution factors at the end of  $t$th iteration step when $b_n^{(t)}$ has been flipped can be directly  updated  by
\begin{align}
\alpha\left( \mathbf b^{(t+1)}, n \right) = -\alpha\left( \mathbf b^{(t)}, n \right) > 0
\label{ch2_contributionupdate1}
  \end{align}
and
  \begin{align}
   \alpha\left( \mathbf b^{(t+1)}, m \right) = \alpha\left( \mathbf b^{(t)}, m \right) - 4 b_{m}^{(t)}b_{n}^{(t)} \mathbf y_m^\top \mathbf y_n 
\label{ch2_contributionupdate2}
\end{align}
for every $m \neq n$. Given the data correlation matrix $\mathbf Y^\top \mathbf Y$, updating the contributions by \eqref{ch2_contributionupdate1} and \eqref{ch2_contributionupdate2}  
costs only   $\mathcal O(N)$.
  For ease in reference, complete code for L1-BF, $K=1$, is provided in Fig. \ref{ch2_algo1}.

\noindent \emph{Convergence Characterization} 

The termination condition of the BF iterations (convergence) is summarized to 
\begin{align} 
\alpha\left( \mathbf b^{(t)}, n \right) \geq 0 ~\forall n \in \{1, 2, \ldots, N \}.
\label{ch2_terminationcondition}
\end{align}
That is, when \eqref{ch2_terminationcondition} is met, the algorithm returns $\mathbf b_{\mathrm{bf} }= \mathbf b^{(t)}$ and terminates. 
For any initialization point,  \eqref{ch2_terminationcondition} will   be met and the BF iterations will converge in a finite number of steps, since  (i)   the binary quadratic form maximization in \eqref{ch2_quad}  has a finite upper bound and (ii)   at every step of the presented BF iterations the quadratic value increases. 
Thus, the proposed iterations converge globally\footnote{We say that a sequence converges \emph{globally} if it convergences for any initialization.} both in argument and in value.

\noindent \emph{Characterization of     Point of Converngence and Relationship to   Fixed Points of \cite{Kwak2008, Nie2011}}

The last point of the generated sequence, $\mathbf b_{\mathrm{bf}}$, satisfies the termination condition $\alpha \left(\mathbf b_{\mathrm{bf}}, {n} \right)  \geq 0~ \forall n$, which in view of \eqref{ch2_contribution} can be equivalently rewritten as 
$  b_{\mathrm{bf},n} \mathbf y_n^\top\mathbf Y \mathbf b_{\mathrm{bf}} \geq ||\mathbf y_n||_2^2 ~\forall n$.
Therefore,   for every initialization point, the  BF iterations converge   in the  set
\begin{align}
\Omega (\mathbf Y)  = \{\mathbf b \in \{\pm1\}^{N};~~  
      b_n\mathbf y_n^\top\mathbf Y \mathbf b \geq ||\mathbf y_n||_2^2 ~ \forall n  \}.
\label{ch2_sbfset2} 
\end{align}
By \eqref{ch2_XtoY}, $\mathbf Y^\top \mathbf Y = \mathbf X^\top \mathbf X$ and, thus, $\Omega(\mathbf Y) = \Omega (\mathbf X)$. Importantly, the following proposition holds true.
\begin{myprop}
Every optimal solution to the binary-quadratic maximization in \eqref{ch2_quad2}  belongs to $\Omega (\mathbf Y)$.
\label{ch2_conditionaloptimality}
\end{myprop}

\noindent{\emph{Proof:}} Let $\mathbf b$ be a solution to \eqref{ch2_quad2} outside $\Omega (\mathbf Y)$. Then, there exists at least one entry in $\mathbf b$, say the $n$th entry, for which the contribution to the objective value is negative; i.e.,
$
\alpha\left( \mathbf b^{(t)}, n \right) <0.
$
Consider now the binary vector $\mathbf b' \in \{\pm 1\}^N$   derived by negation of  the $n$th entry of $\mathbf b$. By \eqref{ch2_metricincrease},   $\left\| \mathbf Y \mathbf b'\right\|_2^2 =  \left\| \mathbf Y \mathbf b\right\|_2^2 - 2| \alpha_{n}| > \left\| \mathbf Y \mathbf b\right\|_2^2$. Hence, $\mathbf b$ is not a maximizer in \eqref{ch2_quad}. 
\hfill $\qed$

By Proposition \ref{ch2_conditionaloptimality}, being an element of $\Omega (\mathbf Y)$ is a necessary condition for a binary vector to be a solution to \eqref{ch2_quad2}.
Since  any point in $\Omega (\mathbf Y)$ is reachable by the BF iterations upon appropriate initialization,   L1-BF may indeed calculate $\mathbf b_{\mathrm{opt}}$ and, through \eqref{ch2_bbftoqbf}, return the optimal L1-PC  solution to \eqref{ch2_l1pca},  $\mathbf q_{L1}$.
 The following Proposition establishes an important relationship between the set of convergence points of the proposed BF iterations, $\Omega (\mathbf Y)$, and the set of convergence points of the fixed-point iterations in  \cite{Kwak2008, Nie2011}, say $\Phi(\mathbf Y)$.

\begin{myprop}
Every element of $\Omega (\mathbf Y)$ lies also in the fixed-point set $\Phi (\mathbf Y)$; that is, $\Omega (\mathbf Y) \subseteq \Phi (\mathbf Y)$.
\label{ch2_relationship} \end{myprop}  

\noindent \emph{Proof:}
Consider an arbitrary $\mathbf b \in \Omega (\mathbf Y)$. Then, 
$
b_n\mathbf y_n^\top\mathbf Y \mathbf b \geq ||\mathbf y_n||_2^2,
$ for all $  n \in \{1, 2, \ldots, N \}$. 
Assume  that $b_n = +1$ for some $n$. Then,   $ \mathbf y_n^\top\mathbf Y \mathbf b \geq \left\| \mathbf y_{n} \right\|_2^2 \geq 0$ and $\textrm{sgn}(\mathbf y_n^\top\mathbf Y \mathbf b) = +1 = b_{n}$. 
Otherwise, $b_n = -1$. Then,  $ \mathbf y_n^\top\mathbf Y \mathbf b \leq -\left\| \mathbf y_{n} \right\|_2^2 \leq 0$ and  $ \textrm{sgn}(\mathbf y_n^\top\mathbf Y \mathbf b) = -1 = b_{n}$. %
Hence, for every vector $\mathbf b\in \Omega (\mathbf Y)$,  $b_{n} =   \mathrm{sgn}(\mathbf y_{n}^\top \mathbf Y \mathbf b)$  for all $n$ and, thus, $\mathbf b \in \Phi (\mathbf Y)$.  %
\hfill $\square$

If we denote by $B(\mathbf Y)$ the set of optimal solutions to \eqref{ch2_quad}, Propositions \ref{ch2_conditionaloptimality} and \ref{ch2_relationship} can be summarized as 
\begin{align}
B(\mathbf Y) \subseteq \Omega(\mathbf Y)  \subseteq \Phi (\mathbf Y) \subseteq  \{ \pm 1 \}^{N}
\label{ch2_subsetrelationship}
\end{align}
or, in terms of set sizes, 
$2 \leq |B (\mathbf Y)| \leq |\Omega (\mathbf Y)| \leq |\Phi(\mathbf Y)| \leq 2^N $. 

\begin{figure}[h]
\centering
                    \includegraphics[width=0.8\linewidth]{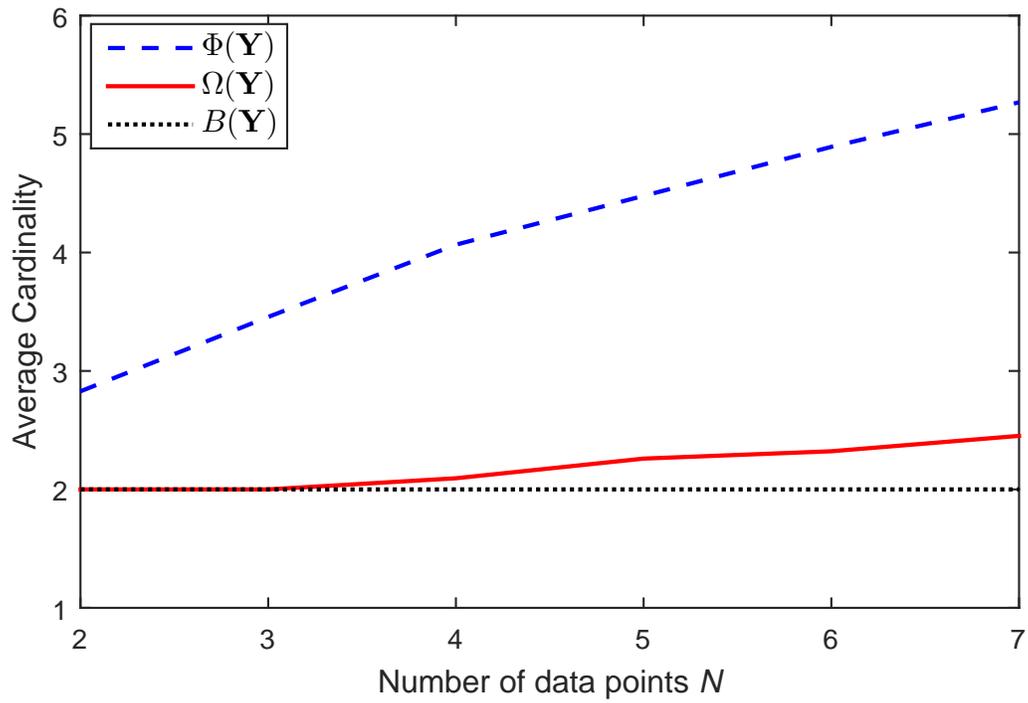}
                    \vspace{0.1cm}
         \caption{Average cardinality of fixed-point set $\Phi(\mathbf Y)$ \cite{Kwak2008,Kwak2009}, bit-flipping convergence set $\Omega(\mathbf Y)$, and set of optimal points $B(\mathbf Y)$ versus  number of data points $N$ ($d=2$).} 
                         \label{ch2_fig:cardinalities}
\end{figure}

For illustration, in Fig. \ref{ch2_fig:cardinalities}, we demonstrate experimentally the relationship between the cardinalities of $\Phi(\mathbf Y)$, $\Omega(\mathbf Y)$, and  $B(\mathbf Y)$. We generate $1000$ independent instances of a full-rank data matrix $\mathbf Y \in \mathbb R^{(d=2) \times N}$ for $N=2,3, \ldots, 7$, with each element of $\mathbf Y$  drawn independently from the Gaussian $\mathcal N( 0, 1)$ distribution. Then, we  plot   the average observed cardinality of $\Phi(\mathbf Y)$, $\Omega(\mathbf Y)$, and  $B(\mathbf Y)$ versus the number of data points $N$. 
We see that as $N$ increases $\Omega(\mathbf Y)$ remains a tight super-set of $B(\mathbf Y)$, while the cardinality of  $\Phi(\mathbf Y)$ (number of possible converging points of the FP iterations) increases at a far higher rate. 
 Fig. \ref{ch2_fig:cardinalities} offers insight why the proposed BF iterations are expected to return the optimal solution to \eqref{ch2_quad} much more often  than the FP iterations of   \cite{Kwak2008, Kwak2009, Nie2011}.

\noindent \emph{Performance Bounds}  
 
 The performance attained by   $\mathbf q_{\mathrm{bf}}$ is formally lower bounded by Proposition \ref{ch2_bounds} whose proof is given in the Appendix.

\begin{myprop} \label{lowerbound}
The performance  of $\mathbf q_{\mathrm{bf}}$ in the metric of \eqref{ch2_l1pca}, calculated upon any initialization, is lower bounded by
$
 \left\| \mathbf X^\top \mathbf q_{\mathrm{bf}}\right\|_1  \geq \left\| \mathbf X\right\|_F. 
$
\label{ch2_bounds}
\end{myprop}  

\begin{figure}[h]
\centering
\includegraphics[width=0.8\linewidth]{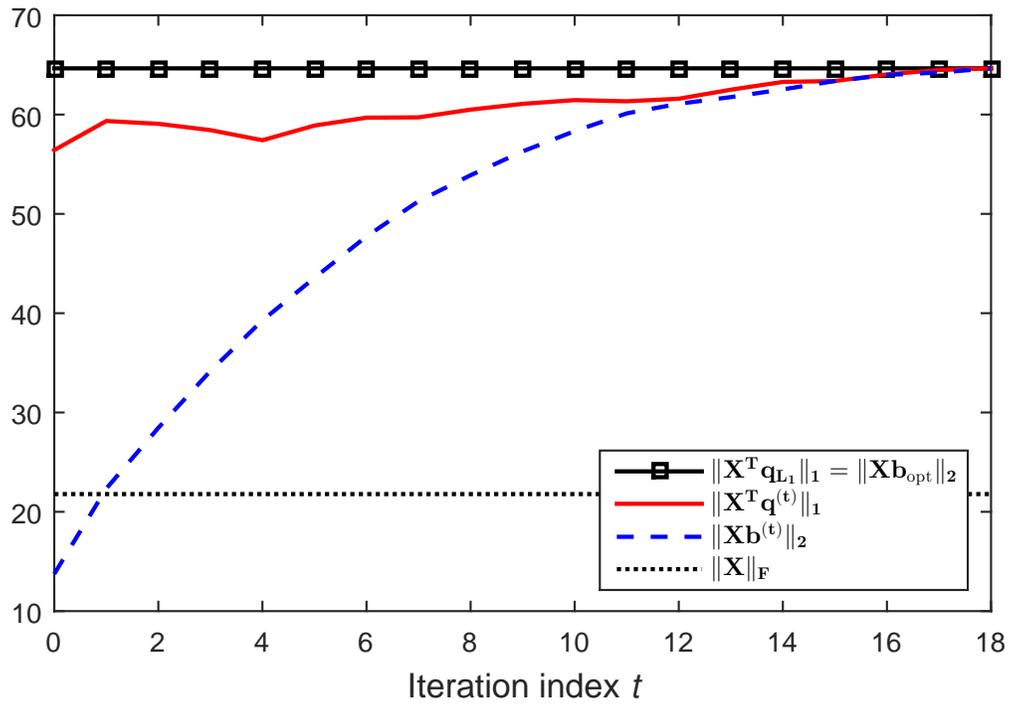}
        \caption{
        	Binary quadratic metric $\| \mathbf X^\top \mathbf b^{(t)}\|_2$ and  L1-PCA metric $\| \mathbf X^\top \mathbf q^{(t)}\|_1$ per iteration. We plot along the  upper bound line  $\| \mathbf X^\top \mathbf q_{L1}\|_1 = \| \mathbf X \mathbf b_{\mathrm{opt}}\|_2$ and lower bound line $\| \mathbf X\|_F$ of Proposition \ref{lowerbound}.}
        \label{ch2_figmetrics}
\end{figure}

Moreover, for the optimal L1-PC $\mathbf q_{L1}$ 
\begin{align}
\left\| \mathbf X^\top \mathbf q_{L1}\right\|_1 &=\left\| \mathbf X \mathbf b_{\mathrm{opt}}\right\|_2    
    = \sqrt{N} \underset{\mathbf z \in \{\pm \frac{1}{\sqrt{N}} \}^{N}}{\mathrm{max}}\left\| \mathbf X \mathbf z \right\|_2   \leq \sqrt{N}   \underset{\mathbf z \in \mathbb R^{N};~\left\| \mathbf z\right\|_2=1}{\mathrm{max}}\left\| \mathbf X \mathbf z \right\|_2 = \sqrt{N} \sigma_{\max}
\label{ch2_bound2}
\end{align}
where $\sigma_{\max}$ is the maximum singular value of the data matrix $\mathbf X$.
Then, the loss with respect to the L1-PCA metric in \eqref{ch2_l1pca} experienced by  $\mathbf q_{\mathrm{bf}}$ is  bounded by
\begin{align}
\left\| \mathbf X^\top \mathbf q_{L1}\right\|_1 -\left\| \mathbf X^\top \mathbf q_{\mathrm{bf}}\right\|_1 \leq \sqrt{N} \sigma_{\max} -\left\| \mathbf X\right\|_F.
\label{bounds}
\end{align}
In Fig. \ref{ch2_figmetrics}, we execute the proposed BF iterations  on an arbitrary data matrix $\mathbf X \in \mathbb R^{(d=4) \times (N=32)}$ (elements drawn independently from $\mathcal N(0,1)$) and plot the binary quadratic metric $\| \mathbf X \mathbf b^{(t)}\|_2$ and  L1-PCA metric $\| \mathbf X^\top \mathbf q^{(t)}\|_1$ per iteration $t$. 
For reference,  we also plot the optimal line $\| \mathbf X^\top \mathbf q_{L1}\|_1 = \| \mathbf X \mathbf b_{\mathrm{opt}}\|_2$ and  lower bound $\| \mathbf X\|_F$. 
Fig. \ref{ch2_figmetrics} offers a vivid numerical illustration of Proposition \ref{lowerbound} and eq. \eqref{bounds}.

\noindent \emph{Initialization}

\begin{figure}[h]
	\centering
		\includegraphics[width=0.8 \linewidth]{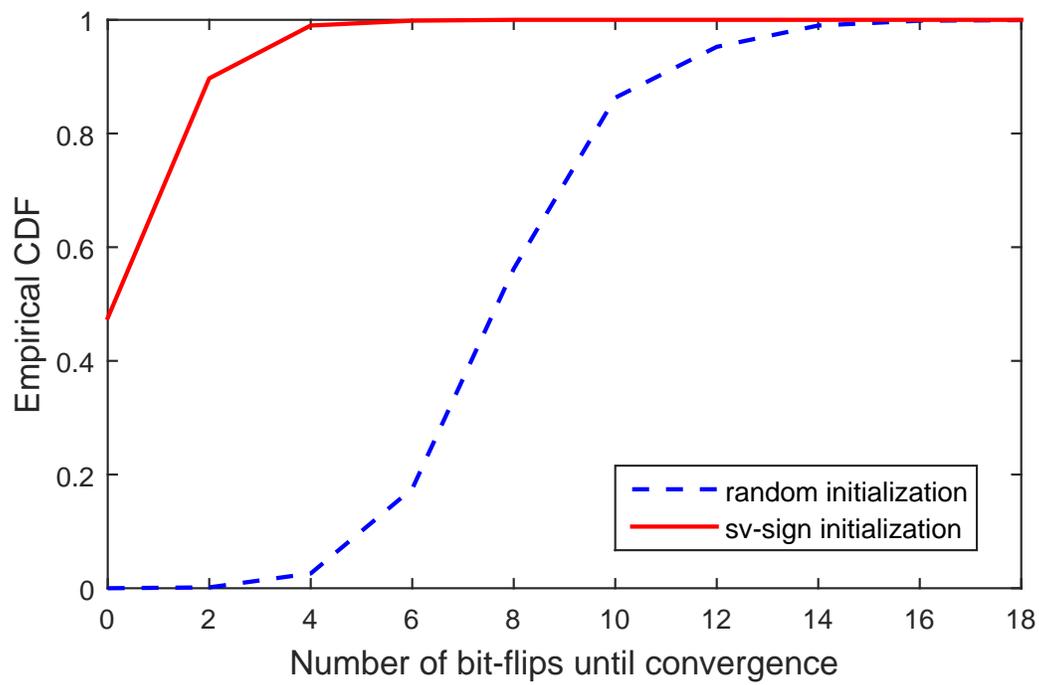}
	\caption{Empirical CDF of number of bit-flips needed until convergence for sv-sign and random initialization ($D=3$, $N=20$, $1000$ experiments).}
		\label{fig:initnbf}
\end{figure}

Next, we  consider   initializations that may both expedite convergence and  offer  superior L1-PC approximations.  In the trivial $d=1$ case,   $\mathbf Y =   \sigma \mathbf v^\top$ for some    $\sigma \in \mathbb R_{+}$  and $\mathbf v \in \mathbb R^{N \times 1}$, $\left\| \mathbf v\right\|_2 =1 $, and
$\mathbf b_{\mathrm{opt}} 
=  {\arg \max}_{{\mathbf b \in \mathbb \{ \pm 1\}^{N}}} \left\| \mathbf Y \mathbf b\right\|_2^2 =  {\arg \max}_{ {\mathbf b \in \mathbb \{ \pm 1\}^{N}}} | \mathbf v^\top \mathbf b|_2^2 =
 \pm \mathrm{sgn}(\mathbf v) 
$.
Motivated by this special case, we  initialize the BF iterations
to the sign of the right singular vector of $\mathbf Y$ that corresponds to the highest singular value. 
We call this initialization choice \emph{sv-sign initialization}. Evidently, by the definition of $\mathbf Y$ in \eqref{ch2_defY},  
\begin{align}
\mathbf b^{(1)} = \mathrm{sgn}([\mathbf Y^\top]_{:,1}).
\label{ch2_vopt}
\end{align}
Experimental studies that compare the efficiency of  the proposed sv-sign initialization with that of  equiprobable random initializations ($\mathbf b^{(1)}$ takes any value in $\{ \pm 1\}^N$ with probability $2^{-N}$) show that sv-sign initialization   not only leads to superior L1-PC approximations with respect to \eqref{ch2_l1pca}, but also  reduces significantly  the actual execution time of the proposed  algorithm by reducing the number of bit-flips needed for convergence. As an illustration, we generate  $1000$ realizations of the  data matrix $\mathbf X \in \mathbb R^{3 \times 20}$ with entries drawn independently from $\mathcal N(0,1)$. On each realization we run  bit-flipping iterations with both sv-sign and equiprobable binary initialization. We observe that $81.6\%$ of the time the L1-PC obtained with sv-sign initialization attains value in the metric of \eqref{ch2_l1pca} greater than or equal to  L1-PC obtained by random initialization. 
Also,  in Fig. \ref{fig:initnbf} we plot the empirical CDF of the number of bit-flips needed until convergence for the two initializations. We observe that $50\%$ of the time   sv-sign initialization is already in the convergence set $\Omega(\mathbf X)$ and no bit-flips take place. Moreover, with sv-sign initialization no more than $6$ bit-flips are needed for convergence with empirical probability $1$. On the contrary, random initialization may need with non-zero probability up to $16$ bit-flips for convergence.

\noindent \emph{Complexity Analysis}  

Prior to   bit-flipping iterations, L1-BF calculates $\mathbf Y$ in \eqref{ch2_defY} and chooses the initial  binary vector $\mathbf b^{(1)}  $ with  complexity $\mathcal O(ND \mathrm{min}\{ N,D \})$ (SVD). Then, the algorithm  calculates   and stores  the correlation  matrix $\mathbf Y^\top \mathbf Y \in \mathbb R^{N \times N}$  with complexity $\mathcal O(N^2 d)$ and,  given  $ \mathbf b^{(1)}$ and $\mathbf Y^\top \mathbf Y$,  evaluates by \eqref{ch2_contribution} the initial contribution factors
$\alpha \left(\mathbf b^{(1)}, {n} \right)$ for all $n \in \{1, 2, \ldots, N \}$ with cost  $\mathcal{O}(N^2)$. 
Therefore,  initialization has total cost  $\mathcal O (ND \mathrm{min}\{N,D\} + N^2 (d+1))$.
At iteration  $t \geq 1$, finding   $n$  by \eqref{ch2_update2} costs $\mathcal{O}(N)$. 
In  case  $\alpha \left(\mathbf b^{(t)}, {n} \right) > 0$,    repetition of the maximization over the entire index set $\{1, 2, \ldots, N \}$    costs an additional $\mathcal{O}(N)$.
Therefore, each bit flip costs $\mathcal O( N)$.
 Denoting    by $ M$ the number of bit flips   until the termination criterion is met, we find that the computational cost  for calculating  $\mathbf b_{\mathrm{bf}}$ is 
   $\mathcal{O}(ND \mathrm{min}\{ N,D\} + N^2 (d+1) +  MN)$.
 
   Finally, taking into account the computation of $\mathbf q_{\mathrm{bf}}$ from $\mathbf  b_{\mathrm{bf}}$, the proposed  L1-BF algorithm costs  $\mathcal{O}(ND $ $ \mathrm{min}\{ N,D \}$ $ + N^2 (d+1) + (M+D)N)$. 
   According to our experiments, 
    $M$ is with empirical probability $1$ upper bounded by $N$; that is,  no more than $N$ bits need to be flipped to reach $\mathbf b_{\mathrm{bf}}$.\footnote{In practice, a brute-force termination  $M \leq N$  can be imposed  to the algorithm.}    
   Therefore,  the total complexity of the proposed algorithm becomes $\mathcal{O}(ND \mathrm{min}\{ N,D\} + N^2 (d+2) + ND)$. Keeping only the dominant complexity terms, L1-BF has complexity $\mathcal O(ND \mathrm{min}\{N,D\} + N^2d)$, i.e.  quadratic in $N$ and  either linear, or quadratic in $D$ (depending on the cost of the initial SVD). Thus, the proposed L1-BF algorithm has cost comparable to that of standard PCA (SVD).
   A summary of the calculated complexity  is provided in Table \ref{ch2_table:complexity}.

\noindent \emph{Multiple Initializations}
 
For further  L1-PCA metric improvement, we may also run BF  on  $L$ distinct initialization points, 
$  \mathbf b^{(1)}_{1} =\mathrm{sgn}([\mathbf Y^\top]_{:,1})$  and    $\mathbf b^{(1)}_{l} =\mathrm{sgn}(\mathbf Y^\top \mathbf a_{l})$, with $\mathbf a_{l} \sim \mathcal N(\mathbf 0_{d}, \mathbf I_{d})$,  $l = 2, \ldots, L$, to obtain  $L$ corresponding convergence points $\mathbf b_{\mathrm{bf},1}, \mathbf b_{\mathrm{bf},2}, \ldots, \mathbf b_{\mathrm{bf},L}$. 
Then,  if $\mathbf q_{\mathrm{bf},l} =  \mathbf X \mathbf b_{\mathrm{bf},l}\left\| \mathbf X \mathbf b_{\mathrm{bf},l}  \right\|_2^{-1}$, $l=1, \ldots, N$, we return
$
\mathbf q_{\mathrm{bf}}^{(L)} =    {\mathrm{argmax}}_{\mathbf q \in \{ \mathbf q_{\mathrm{bf},1},   \ldots,  \mathbf q_{\mathrm{bf},L} \}} \left\| \mathbf X^\top \mathbf q \right\|_1.
$
Certainly, multiple initializations will increase the complexity of  the algorithm  by a constant factor $L$.

\subsection{Calculation of  $K>1$ L1-Principal Components}
\label{multiple_components}

In contrast to L2-PCA,   L1-PCA in \eqref{ch2_l1pca} is not a scalable problem \cite{Panos2}.
 Therefore,  successive-nullspace-projection approaches  \cite{Kwak2008,McCoy2011} fail to return optimal L1-PC bases. Optimal L1-PCA  demands joint computation of all %
$K$ principal components of $\mathbf Y$, a procedure that increases exponentially in $K$ the computational complexity of optimal algorithms. %
In this section, we generalize the BF algorithm presented in Section \ref{single_component} to  calculate $K>1$   L1-PCs   of any given data matrix $\mathbf X \in \mathbb R^{D\times N}$ of rank $d \geq K$.

 Given the reduced-size, full-row-rank data matrix $\mathbf Y \in \mathbb R^{d \times N}$ (see Proposition \ref{ch2_fromXtoY}), the proposed algorithm first attempts to approximate the solution to \eqref{ch2_nucnormmet} $\mathbf B_{\mathrm{opt}}$.
To do so, it begins from an   initial matrix $ \mathbf B^{(1)}$ and employs   bit-flipping   iterations to reach an approximation to $\mathbf B_{\mathrm{opt}}$, say $\mathbf B_{\mathrm{bf}}$. 
Finally, per \eqref{ch2_BhtoQh} the algorithm returns
  \begin{align}
  \mathbf Q_{\mathrm{bf}} = U \left( \mathbf X\mathbf B_{\mathrm{bf}} \right).
  \label{ch2_BhtoQh}
   \end{align}
Evidently,  if $\mathbf B_{\mathrm{bf}}$ is an exact solution to  \eqref{ch2_nucnorm}, then $\mathbf Q_{\mathrm{bf}}$ is an exact solution to the L1-PCA problem in \eqref{ch2_l1pca}.

\noindent \emph{Calculation of $\mathbf B_{\mathrm{bf}}$ via Bit-flipping Iterations}  

The algorithm initializes at a binary matrix ${\mathbf B^{(1)}} $ and employs BF iterations  to  generate  a sequence of binary matrices $\mathbf B^{(t)}$, $t=2,3,\ldots$, such that  each matrix $\mathbf B^{(t)}$ attains the highest possible value in the maximization metric of  \eqref{ch2_nucnormmet} given that it differs from $\mathbf B^{(t-1)}$  in exactly one entry.
Similar to the $K=1$ case,  the indices of  bits that have not been flipped throughout the iterations  are kept in a memory index-set $\mathcal L$  initialized at $\{1, 2, \ldots, NK \}$ (the $(n,k)$th bit of the binary matrix argument in \eqref{ch2_nucnormmet} has corresponding single-integer index $(k-1)N+n \in \{1, 2, \ldots, NK \}$). 
 
Mathematically, at the $t$th iteration step,  $t \geq 1$, we search for the single bit in the current binary argument $\mathbf B^{(t)}$ whose index  is not in $\mathcal L$ and   its flipping  increases the nuclear-norm metric $\| \mathbf Y \mathbf B^{(t)}\|_*$ the most. We notice that if we flip the $(n,k)$th bit of $\mathbf B^{(t)}$ setting $\mathbf B^{(t+1)} = \mathbf B^{(t)} - 2 B_{n,k}^{(t)}\mathbf e_{n,N} \mathbf e_{k,K}^\top$, then  
\begin{align}
 \mathbf Y \mathbf B^{(t+1)} = \mathbf Y \mathbf B^{(t)} - 2B_{n,k}^{(t)}  \mathbf y_{n} \mathbf e_{k,K}^\top.
 \label{ch2_metricincrease2}
\end{align} 
Therefore, at the $t$th iteration step, the algorithm obtains the index pair 
 \begin{align}
(n,k) = \underset{ \begin{smallmatrix}(m,l) \in \{1, 2, \ldots, N \} \times \{1, 2, \ldots, K \} \\ (l-1)N+m \in \mathcal L \end{smallmatrix}}{\mathrm{argmax}} \left\| \mathbf Y \mathbf B^{(t)}  - 2 B_{m,l}^{(t)} \mathbf y_{m}\mathbf e_{l,K}^\top \right\|_*.
 \label{ch2_update3}
\end{align} 
 If     $ \left\| \mathbf Y \mathbf B^{(t)}  - 2 B_{n,k}^{(t)} \mathbf y_{n}\mathbf e_{k,K}^\top \right\|_* > \left\| \mathbf Y \mathbf B^{(t)} \right\|_*$,  the algorithm flips $B_{n,k}^{(t)}$ setting  
$ \mathbf B^{(t+1)} = \mathbf B^{(t)} - 2 B_{n,k}^{(t)} \mathbf e_{n,N}\mathbf e_{k,K}^\top$
  and  updates $\mathcal L$ to $\mathcal L \setminus \{ (k-1)N+n\}$.
If, otherwise, $ \left\| \mathbf Y \mathbf B^{(t)}  - 2 B_{n,k}^{(t)} \mathbf y_{n}\mathbf e_{k,K}^\top \right\|_* \leq \left\| \mathbf Y \mathbf B^{(t)} \right\|_*$, the algorithm obtains a new solution $(n,k)$ to
\eqref{ch2_update3} after resetting $\mathcal L$  to $\{1, 2, \ldots, NK \}$. 
If now, for this new pair $(n,k)$,    $ \left\| \mathbf Y \mathbf B^{(t)}  - 2 B_{n,k}^{(t)} \mathbf y_{n}\mathbf e_{k,K}^\top \right\|_* > \left\| \mathbf Y \mathbf B^{(t)} \right\|_*$, the algorithm sets 
 $\mathbf B^{(t+1)} = \mathbf B^{(t)} - 2 B_{n,k}^{(t)} \mathbf e_{n,N}\mathbf e_{k,K}^\top$   and  updates   $\mathcal L =  \mathcal L \setminus \{ (k-1)N+n\}$.
 Otherwise, the iterations terminate and the algorithm returns $\mathbf B_{\mathrm{bf}} = \mathbf B^{(t)}$ and $\mathbf Q_{\mathrm{bf}} =  \hat{\mathbf U} \hat{\mathbf V}^\top
$,  where $\mathbf X\mathbf B_{\mathrm{bf}} \overset{\mathrm{ SVD}}{=} \hat{\mathbf U}_{D \times K} \hat{\mathbf \Sigma}_{K \times K}\hat{\mathbf V}_{K \times K}^\top$.  
 
We note that to solve \eqref{ch2_update3} exhaustively, one has to calculate $ \left\| \mathbf Y \mathbf B^{(t)}  - 2 B_{m,l}^{(t)} \mathbf y_{m}\mathbf e_{l,K}^\top \right\|_*$  for all $(m,l)  \in \{1, 2, \ldots, N \} \times \{1, 2, \ldots, K \}$ such that $ (l-1)N+m \in \mathcal L$. 
In worst case, $\mathcal L = \{1, 2, \ldots, NK \}$ and this demands $NK$ independent singular-value/nuclear-norm calculations. 
A simplistic, but computationally inefficient, approach would be perform an SVD on $ \mathbf Y \mathbf B^{(t)}  - 2 B_{m,l}^{(t)} \mathbf y_{m}\mathbf e_{l,K}^\top$ from scratch with cost $\mathcal O(dK^2)$. This would yield a worst case total cost of $\mathcal O(NdK^3)$ to find $(n,k)$ in \eqref{ch2_update3}. Below, we present an alternative method to solve  \eqref{ch2_update3} with cost $\mathcal O(dK^2 + N(K^3 + dK))$.

\noindent \emph{Reduced-cost Nuclear-norm Evaluations for Optimal Bit-flipping}  
 
At the beginning of the $t$th iteration step, we perform the singular-value decomposition
 \begin{align}
 \mathbf Y \mathbf B^{(t)}  \overset{\mathrm{ SVD}}{=} \mathbf U_{d \times K}^{(t)} \mathbf S_{K \times K}^{(t)} \mathbf V_{K \times K}^{{(t)}^\top}
 \end{align}
  with cost $\mathcal O(dK^2)$. 
  Then, for any $(m,l) \in \{1, 2, \ldots, N \}\times\{1, 2, \ldots, K \} $, 
   the singular values of  
$
   \mathbf Y \mathbf B^{(t)}  - 2 B_{m,l}^{(t)} \mathbf y_{m}\mathbf e_{l,K}^\top
$
   are the  square roots of the eigenvalues of $(\mathbf Y \mathbf B^{(t)}  - 2 B_{m,l}^{(t)} \mathbf y_{m}\mathbf e_{l,K}^\top)^\top (\mathbf Y \mathbf B^{(t)}  - 2 B_{m,l}^{(t)} \mathbf y_{m}\mathbf e_{l,K}^\top) $ (found with constant cost), which due to rotation invariance are equal to the eigenvalues of 
$
    \mathbf A =  \mathbf V^{{(t)}^\top} (\mathbf Y \mathbf B^{(t)}  - 2 B_{m,l}^{(t)} \mathbf y_{m}\mathbf e_{l,K}^\top)^\top (\mathbf Y \mathbf B^{(t)}  - 2 B_{m,l}^{(t)} \mathbf y_{m}\mathbf e_{l,K}^\top) \mathbf V^{(t)}.
 $
 Defining 
$\mathbf W = [[\mathbf V^{(t)}]_{l,:}^\top, - 2 B_{m,l}^{(t)}  \mathbf S^{{(t)}^\top}  \mathbf U^{{(t)}^\top}   \mathbf y_{m}] \in \mathbb R^{K \times 2}$ and the eigenvalue decomposition (EVD) 
$\mathbf Q  \mathbf D \mathbf Q^\top \overset{\text{EVD}}{=}  \| \mathbf y_{m}\|_2^2 \mathbf e_{1,2}\mathbf e_{1,2}^\top + [\mathbf e_{2,2}, \mathbf e_{1,2}]$ where $\mathbf D = \mathrm{diag}( [d_1,~d_2]^\top)$,
 it is easy to show that  
 \begin{align}
\mathbf A =  \mathbf S^{{(t)}^\top}\mathbf S^{{(t)}}  + d_1 \mathbf W \mathbf q_{1}(\mathbf W \mathbf q_{1})^\top  + d_2 \mathbf W \mathbf q_{2}(\mathbf W \mathbf q_{2})^\top  .
\end{align}
Consider now the eigenvalue decomposition
\begin{align}
  \mathbf Z_{K \times K} \mathbf P \mathbf Z^\top \overset{\mathrm{EVD}}{= }  \mathbf S^{{(t)}^\top}\mathbf S^{{(t)}} + d_1 \mathbf W \mathbf q_{1}(\mathbf W \mathbf q_{1})^\top.
  \end{align}
   Then, again by eigenvalue permutation invariance, the singular values of $ \mathbf Y \mathbf B^{(t)}  - 2 B_{m,l}^{(t)} \mathbf y_{m}\mathbf e_{l,K}^\top$ are equal to the square roots of the eigenvalues of 
   \begin{align}
 \mathbf Z^\top \mathbf A  \mathbf Z = \mathbf P  +  d_2 \mathbf Z^\top \mathbf W \mathbf q_{2}(\mathbf Z^\top \mathbf W \mathbf q_{2})^\top .
 \end{align}
 Notice that matrix ${\mathbf S^{(t)}}^\top {\mathbf U^{(t)}}^\top$ needed for the calculation of $\mathbf W$  is constant for all candidate-bit evaluations of the $t$th iteration and,  given the SVD of $\mathbf Y \mathbf B^{(t)}$, can be found with cost $\mathcal O(K^2)$. Thus, this computational cost is absorbed in the SVD of $\mathbf Y \mathbf B^{(t)}$. Then, for bit-flip  examination of all $K$ bits of the $m$th row of $\mathbf B^{(t)}$ it suffices to calculate   the quantities $\mathbf Q$, $\mathbf d$, and ${\mathbf  S^{(t)}}^\top {\mathbf U^{(t)}}^\top \mathbf y_{m}$, appearing in $ \mathbf S^{{(t)}^\top}\mathbf S^{{(t)}}  + d_1 \mathbf W \mathbf q_{1}(\mathbf W \mathbf q_{1})^\top$ and $ d_2 \mathbf W \mathbf q_{2}$, just once with cost $\mathcal O(dK)$. 
   Employing the algorithm proposed in \cite{GolubSVD} for the EVD of a diagonal positive semidefinite matrix perturbated by a rank-1 symmetric matrix,   EVD  of $  \mathbf S^{{(t)}^\top}\mathbf S^{{(t)}}  + d_1 \mathbf W \mathbf q_{1}(\mathbf W \mathbf q_{1})^\top$ (i.e.,  calculation of $\mathbf Z$ and $\mathbf P$) costs $\mathcal O(K^2)$. Then, defining $d_2 \mathbf Z^\top \mathbf W \mathbf q_{2}(\mathbf Z^\top \mathbf W \mathbf q_{2})^\top$ and finding the eigenvalues of $\mathbf P + d_2 \mathbf Z^\top \mathbf W \mathbf q_{2}(\mathbf Z^\top \mathbf W \mathbf q_{2})^\top$ (again per \cite{GolubSVD}) costs an additional $\mathcal O(K^2)$. Thus, including the initial SVD, the total cost for bit-flipping examination of all bits in $\mathbf B^{(t)}$ (worst case scenario) is  $\mathcal O(dK^2) + N \mathcal O(dK) + NK \mathcal O(K^2) \equiv \mathcal O(dK^2 + N(K^3 + dK) ).$

 \noindent \emph{Termination Guarantee}  

 Similarly to the  $K=1$ case, the termination condition 
\begin{align}
& \left\| \mathbf Y \mathbf B^{(t)}  - 2 B_{m,l}^{(t)} \mathbf y_{m} \mathbf e_{l,K}^\top \right\|_* \leq \left\| \mathbf Y \mathbf B^{(t)} \right\|_* ~~ \forall (m,l) \in \{1, 2, \ldots, N \} \times \{1, 2, \ldots, K \}
\end{align}
is guaranteed to be met after a finite number of iterations  because  (i) binary nuclear-norm maximization in \eqref{ch2_nucnorm} %
has finite upper bound and (ii) at every step of the BF iterations  the nuclear-norm value increases.

\noindent \emph{Initialization}  

 Generalizing the initialization idea for the $K=1$ case, we choose $\mathbf B^{(1)} = \mathrm{sgn}([\mathbf Y^\top]_{:,1}) \mathbf 1_{K}^\top $ where   $\mathbf 1_{K}$ is the all-one vector of length $K$.
Again, our motivation here lies in the fact that for $d=1$ and any $K$, $\mathbf B_{\mathrm{opt}} = \mathrm{sgn}(\mathbf v_{\mathrm{opt}}) \mathbf 1_K^\top$. For this particular initialization, 
$
\left\| \mathbf Y \mathbf B^{(1)}\right\|_* = K\left\| \mathbf Y  \mathrm{sgn}([\mathbf Y^\top]_{:,1}) \right\|_2.
\label{ch2_multK_init_metric}
$  
Pseudocode of the proposed L1-BF algorithm for $K \geq 1$  is offered in Fig. \ref{ch2_algo2}.
Of course,  for $K=1$  the  L1-PCs generated by the algorithms of Fig. \ref{ch2_algo1} and Fig. \ref{ch2_algo2} coincide.

\noindent \emph{Complexity Analysis}  

 Before the BF iterations, the algorithm expends $\mathcal O(ND \mathrm{min}\{N,D\} )$ to calculate $\mathbf Y \in \mathbb R^{d \times N}$ through SVD of $\mathbf X \in \mathbb R^{D \times N}$. By the same SVD, we calculate also the initial binary matrix.
Then, for the proposed initialization $\mathbf Y \mathbf B^{(1)}$ and $\left\| \mathbf Y \mathbf B^{(1)} \right\|_*$ are calculated in the beginning of the first iteration with complexity $\mathcal O(NdK)$.
To find a solution to \eqref{ch2_update3},  we incur 
worst case cost $\mathcal O(dK^2 + N(K^3 + dK) )$.
Setting the maximum number of BF iterations run by the algorithm  to $NK$, the total complexity to calculate $\mathbf B_{\mathrm{bf}}$ is in worst case $\mathcal O (ND \mathrm{min} \{ N,D\} + NdK^3 + N^2(K^4 + dK^2))$.  
Then, to calculate $\mathbf Q_{\mathrm{bf}}$ from $\mathbf B_{\mathrm{bf}}$ costs an additional $\mathcal O(ND \mathrm{min}\{N,D \} + NDK)$.
Therefore, the total complexity of the proposed algorithm  is  $\mathcal O (ND \mathrm{min} \{ N,D\} + N^2(K^4 + dK^2) + NdK^3 )$; that is, quadratic complexity in $N$ and at most quadratic complexity in $D$. Notice that for $K=1$ the asymptotic complexity of the algorithm becomes $\mathcal O(ND \mathrm{min} (N,D) + N^2d)$, i.e. equal to that of the algorithmic operations of Fig. \ref{ch2_algo1}.
A summary of the calculated complexity  is provided in Table \ref{ch2_table:complexity2}.

\noindent \emph{Multiple Initializations}  

Similar to the $K=1$ case, we can further increase  performance of the proposed L1-PCA scheme by running  BF iterations  on  $L$ distinct initialization matrices $\mathbf B^{(1)}_{1}, \mathbf B^{(1)}_{2}, \ldots, \mathbf B^{(1)}_{L}$. 
We choose $\mathbf B^{(1)}_{1} = \mathrm{sgn}([\mathbf Y^\top]_{:,1}) \mathbf 1_{K}^\top$ (sv-sign initialization) and for $l \in \{2,3, \ldots, L\}$ we initialize randomly to $\mathbf B^{(1)}_{l} = \mathrm{sgn} ( \mathbf a_l) \mathbf 1_{K}^\top$,  $\mathbf a_{l} \sim \mathcal N (\mathbf 0_N, \mathbf I_{N})$, 
 to   obtain the $L$ corresponding convergence points $\mathbf B_{\mathrm{bf},1}, \mathbf B_{\mathrm{bf},2}, \ldots, \mathbf B_{\mathrm{bf},L}$ 
and, through  \eqref{ch2_BhtoQh}, the associated  L1-PC bases   $  \mathbf Q_{\mathrm{bf},1}, \mathbf Q_{\mathrm{bf},2}, \ldots,  \mathbf Q_{\mathrm{bf},L} $. Then, we return 
$
\mathbf Q_{\mathrm{bf}}^{(L)} =  {\mathrm{argmax}}_{\mathbf Q \in \{ \mathbf Q_{\mathrm{bf},1}, \mathbf Q_{\mathrm{bf},2}, \ldots,  \mathbf Q_{\mathrm{bf},L} \}}\left\| \mathbf X^\top \mathbf Q \right\|_1.
$

\section{Experimental Studies}

\label{ch2_sec6}

\subsection{Comparison of L1-BF with  Current State-of-the-art Algorithms}

\begin{figure}[h]
\centering
\begin{subfigure}[]{0.8\linewidth}
\centering
\caption{\vspace{-0.30cm}}
\includegraphics[width= \linewidth]{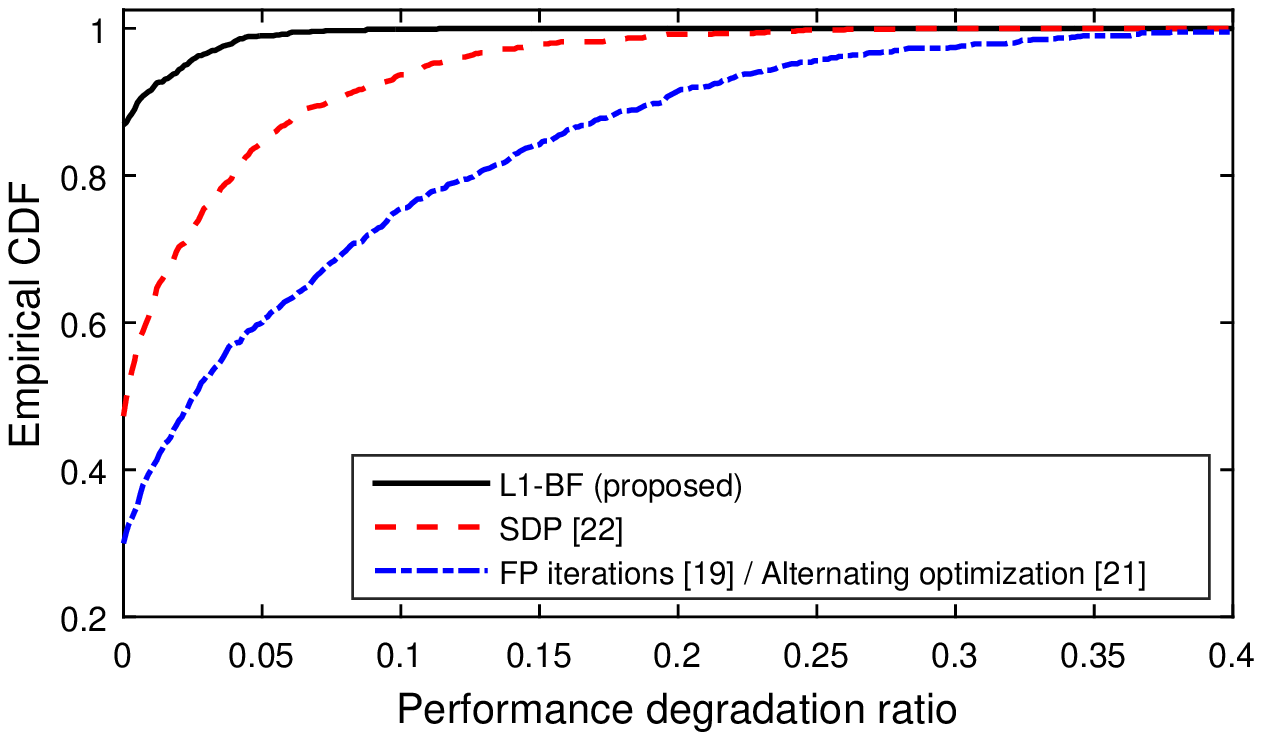}
\label{ch2_fig:comp_soa_K1}
\end{subfigure} \\[0.3cm]
\begin{subfigure}[]{0.8\linewidth}
\centering
\caption{\vspace{-0.30cm}}
\includegraphics[width= \linewidth]{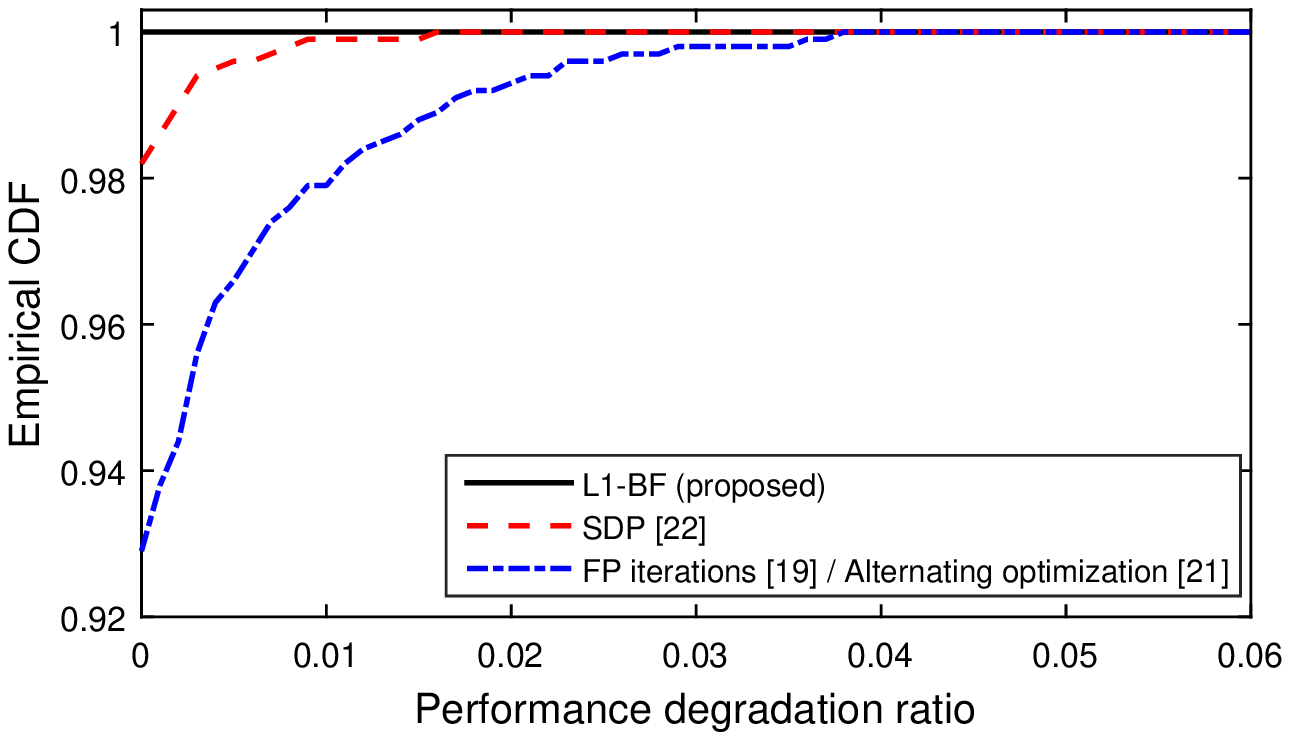}
\label{ch2_fig:comp_soa_K1_2}
\end{subfigure}
        \caption{Empirical CDF  of  $\Delta( \mathbf q_{\mathrm{bf}}; \mathbf X )$, $\Delta (\mathbf q_{\mathrm{fp}}; \mathbf X )$, and $\Delta (\mathbf q_{\mathrm{sdp}}; \mathbf X )$ ($D=4$, $N=16$) for (a) $L=1$ and (b) $L=N=16$ initializations.}
        \label{ch2_fig:compsoa_1}
\end{figure}

We first compare the performance of  L1-BF   with that of   algorithms  proposed in \cite{Kwak2008, Nie2011, McCoy2011}.\footnote{Regarding the SDP approach of \cite{McCoy2011}, the total computational cost  employing   $L$ Gaussian randomization instances  is $\mathcal O(N^{3.5} \mathrm{log}(1/\epsilon) + L(N^2 + DN))$, which is  similar to running the proposed BF iterations  over   $N$ distinct initialization points.}
In our experiment, we generate $1000$ arbitrary matrices of size $(D=d=4) \times (N=16)$ with  entries drawn independently from $\mathcal N(0, 1)$  and calculate the primary ($K=1$) L1-PC  by  L1-BF iterations ($L=1$ initialization), the fixed-point algorithm of \cite{Kwak2008, Nie2011} ($L=1$ initialization),  and the SDP approach in \cite{McCoy2011} ($L=1$ Gaussian randomization).
 We measure  the performance degradation experienced by a unit-length vector $\mathbf q$ on the metric of \eqref{ch2_l1pca} by
  \begin{align}
\Delta(\mathbf q; \mathbf X ) = \frac{\| \mathbf X^\top \mathbf q_{L1}\|_1 - \| \mathbf X^\top \mathbf q\|_1}{\| \mathbf X^\top \mathbf q_{L1}\|_1}
\label{ch2_perfdeg}
\end{align}
 and in Fig. \ref{ch2_fig:comp_soa_K1} we plot the empirical cumulative distribution function (CDF) of  $\Delta( \mathbf q_{\mathrm{bf}}; \mathbf X )$, $\Delta (\mathbf q_{\mathrm{fp}}; \mathbf X )$, and $\Delta (\mathbf q_{\mathrm{sdp}}; \mathbf X )$.
We observe that   $86 \%$ of the time, L1-BF returns the exact, optimal L1-PC of $\mathbf X$. In addition, the performance degradation attained by the proposed algorithm is, with empirical probability $1$, less than $0.09$. 
On the other hand, FP iterations \cite{Kwak2008, Nie2011}  and SDP \cite{McCoy2011} attain optimal performance, with empirical probabilities $0.3$ and $0.5$, respectively. 
The performance degradation for FP \cite{Kwak2008, Nie2011} and SDP \cite{McCoy2011} can be as much as $0.25$ and $0.55$, respectively.

We  repeat the same experiment after increasing the number of initializations for the proposed algorithm and the FP  \cite{Kwak2008, Nie2011} to $L=N=16$  and the number of Gaussian randomizations in SDP \cite{McCoy2011} to $L=N=16$. In   Fig. \ref{ch2_fig:comp_soa_K1_2}, we  plot the corresponding empirical CDFs of  $\Delta( \mathbf q_{\mathrm{bf}}; \mathbf X )$, $\Delta (\mathbf q_{\mathrm{fp}}; \mathbf X )$, and $\Delta (\mathbf q_{\mathrm{sdp}}; \mathbf X )$.
 L1-BF outperforms the two counterparts, returning the exact L1-PC with empirical probability $1$. 

\begin{figure}[h]
\centering
\begin{subfigure}[]{0.8\linewidth}
\centering
\caption{\vspace{-0.30cm}}
\includegraphics[width=\linewidth]{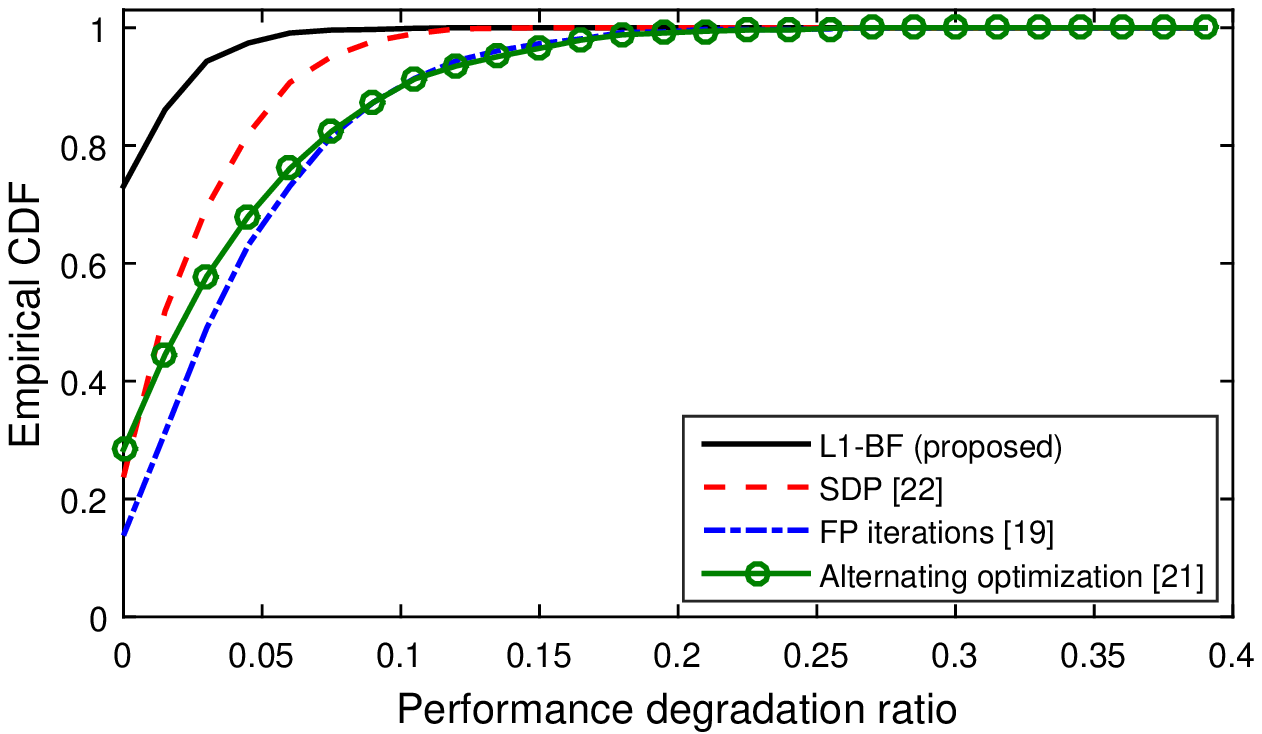}
\label{ch2_fig:compKK}
\end{subfigure} \\[0.3cm]
\begin{subfigure}[]{0.8\linewidth}
\centering
\caption{\vspace{-0.30cm}}
\includegraphics[width=\linewidth]{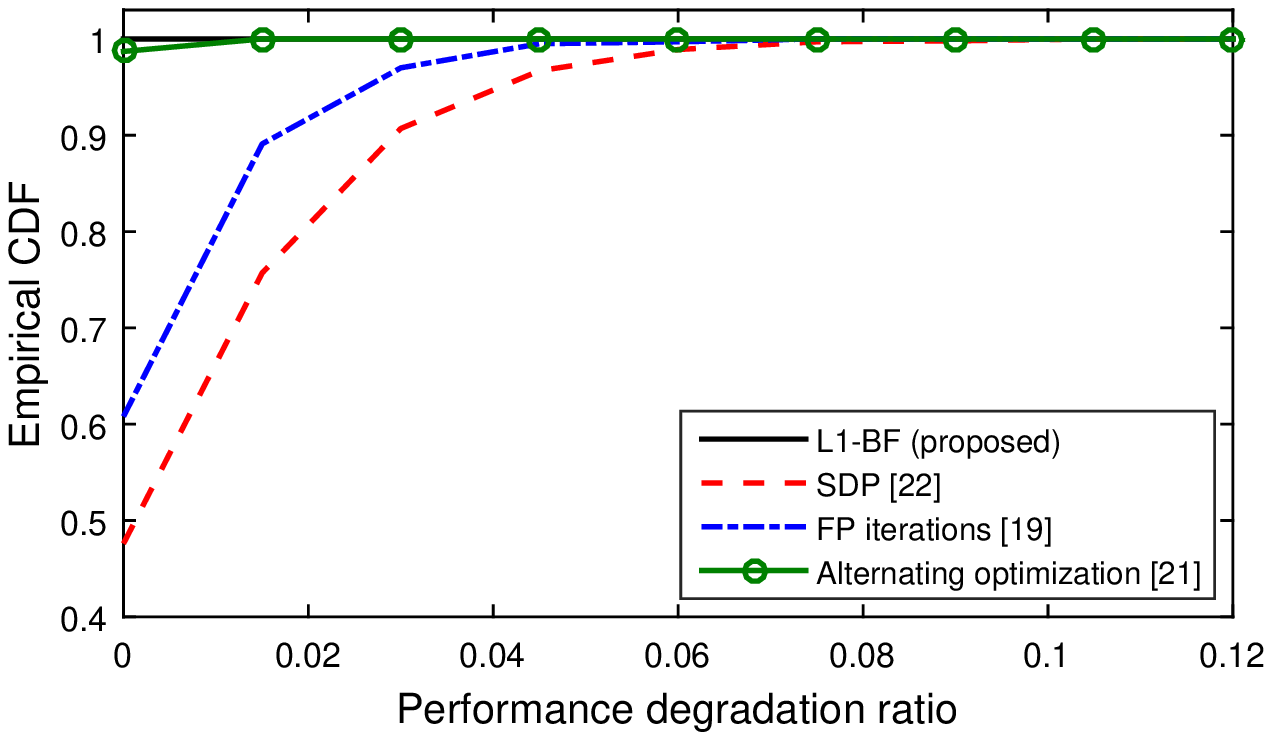}
\label{ch2_fig:compKK2}
\end{subfigure}
        \caption{ Empirical CDF of  $\Delta (\mathbf Q_{\mathrm{ao}}; \mathbf X)$, $\Delta (\mathbf Q_{\mathrm{fp}}; \mathbf X)$,  $\Delta (\mathbf Q_{\mathrm{sdp}}; \mathbf X)$, and $\Delta (\mathbf Q_{\mathrm{bf}}; \mathbf X)$ for (a) $L=1$ and (b) $L=NK=16$ initializations.}
        \label{ch2_fig:composoaKK}
\end{figure}

Next, we examine the performance of the four algorithms (FP iterations with successive nullspace projections \cite{Kwak2008}, alternating optimization \cite{Nie2011}, and SDP with successive nullspace projections \cite{McCoy2011}, and proposed BF iterations) when $K=2$ L1-PCs are sought. 
We generate $1000$ arbitrary data matrices of size $(D=3) \times (N=8)$ (all entries drawn independently from $\mathcal N(0, 1)$) and plot in Fig. \ref{ch2_fig:compKK} the empirical CDF of the corresponding  performance degradation ratios,  $\Delta (\mathbf Q_{\mathrm{fp}}; \mathbf X)$, $\Delta (\mathbf Q_{\mathrm{ao}}; \mathbf X)$, $\Delta (\mathbf Q_{\mathrm{sdp}}; \mathbf X)$, and $\Delta (\mathbf Q_{\mathrm{bf}}; \mathbf X)$. All iterative methods are run on a single initialization point (one Gaussian-randomization instance for SDP).
We observe that   $83 \%$ of the time, L1-BF returns the exact, optimal L1-PCs of $\mathbf X$.
The performance degradation attained by the proposed PCs is, with empirical probability $1$, less than $0.09$. 
On the other hand, the PCs calculated by \cite{ Nie2011, Kwak2008, McCoy2011} attain optimal performance, with empirical probabilities of above $0.31$, $0.05$, and $0.1$, respectively, while
 the performance degradation for the algorithms in \cite{ Nie2011, Kwak2008} and \cite{McCoy2011} can be as much as $30\%$.
 Evidently, the nullspace projections of \cite{  Kwak2008, McCoy2011} that  ``violate" the non-scalability principle of the L1-PCA problem, have significant  impact on their performance. 
Next, we repeat the above experiment with $L=NK=16$ initializations ($L=16$ Gaussian-randomization instances for  SDP). In Fig. \ref{ch2_fig:compKK2}, we plot the new performance degradation ratios for the four algorithms. 
This time, the proposed method returns the exact/optimal L1-PCs of $\mathbf X$ with probability $1$. 
High, but inferior, performance is also attained by the iterative algorithm of \cite{  Nie2011}. On the other hand, the algorithms in \cite{  Kwak2008, McCoy2011}  that follow the greedy approach of solving a sequence of nullspace-projected $K=1$ problem instances experience performance degradation values similar to those of Fig. \ref{ch2_fig:compKK}.

\subsection{Line-fitting Experiment}
\begin{figure}[h]
\centering
\begin{subfigure}[]{0.8\linewidth}
\centering
\caption{\vspace{-0.30cm}}
\includegraphics[width=\linewidth]{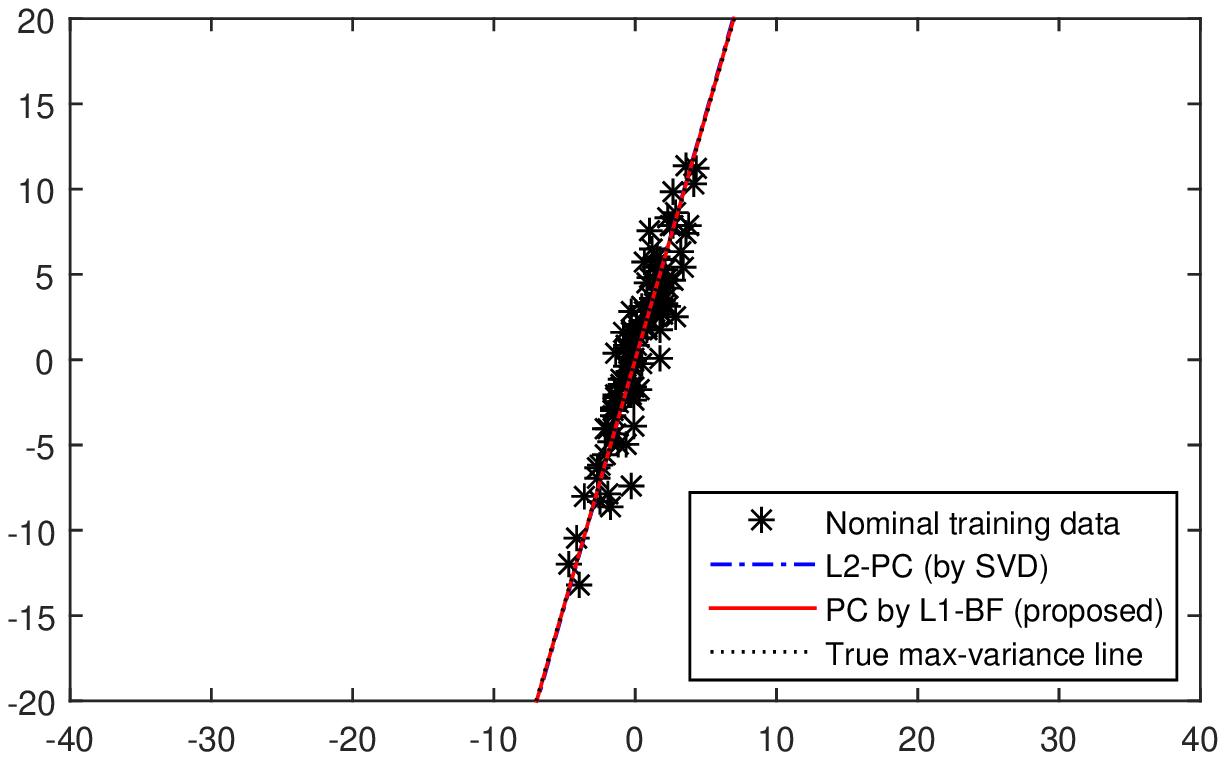}
\label{ch1_fig:linefit1}
\end{subfigure} \\[0.3cm]
\begin{subfigure}[]{0.8\linewidth}
\centering
\caption{\vspace{-0.30cm}}
\includegraphics[width=\linewidth]{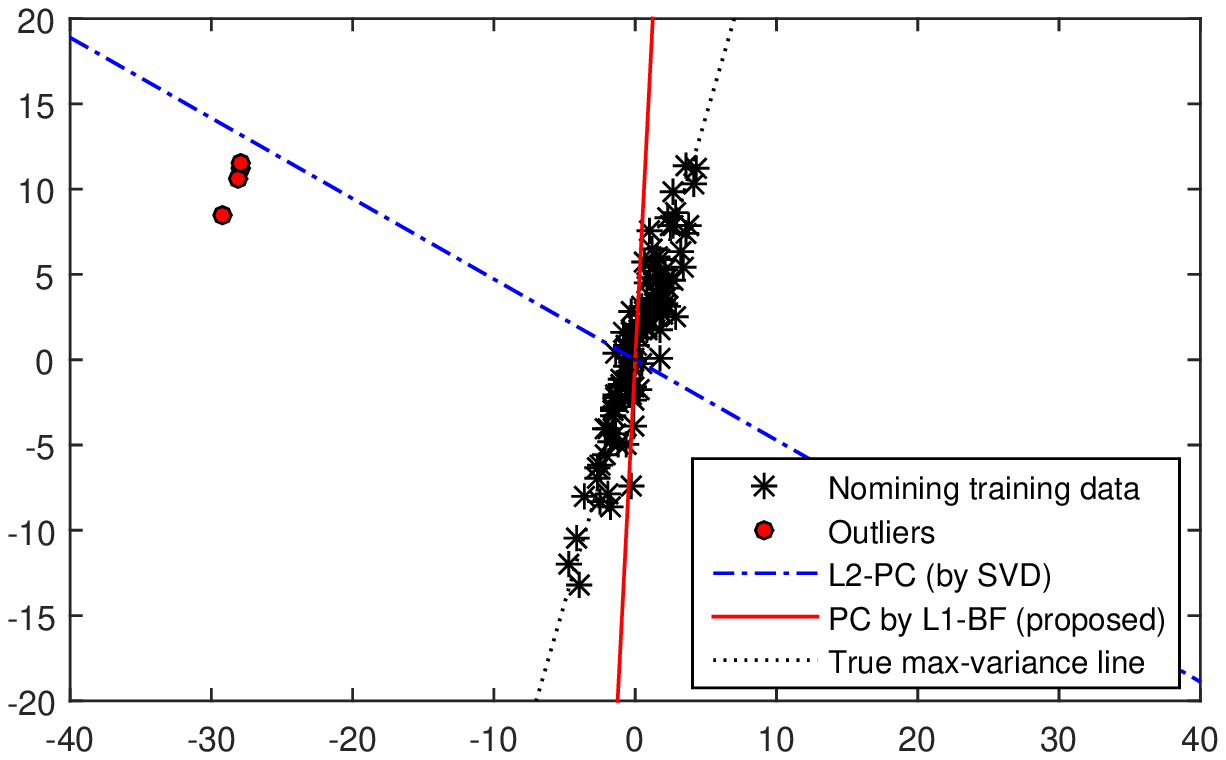}
\label{ch1_fig:linefit2}
\end{subfigure}
        \caption{L2-PC and  L1-PC (by L1-BF) trained over (a) the clean data matrix $\mathbf X \in \mathbb R^{2 \times 100}$ ($100$ nominal data points) and (b) the outlier corrupted data matrix $\mathbf X_{\mathrm{cor}} \in \mathbb R^{2 \times 104}$ (same $100$ nominal data points plus $4$ outliers).} 
        \label{ch1_fig:linefit}
\end{figure}

  \renewcommand{\arraystretch}{0.6}
For visual evaluation of the outlier resistance of the proposed L1-PCs, we consider   $N=100$  $2$-dimensional points drawn from the nominal zero-mean Gaussian distribution  $\mathcal N \left( \mathbf 0_2,
\mathbf R =\begin{bmatrix}
4 & 10 \\ 10 & 29
\end{bmatrix}
\right)$   \renewcommand{\arraystretch}{1}
organized in matrix form 
$\mathbf X_{\mathrm{nom}} = [\mathbf x_{1}, \mathbf x_{2}, \ldots, \mathbf x_{100}]$.
We use the nominal data points to extract an estimate of the true  maximum-variance/minimum-mean-squared-error  subspace (line) of our data distribution 
by means of L2-PCA  (standard SVD method). We do the same  by means of L1-PCA through the proposed L1-BF method. 
In Fig. \ref{ch1_fig:linefit1}, we plot on the $2$-dimensional plain the $100$ training data points in $\mathbf X_{\mathrm{nom}}$ and the lines described by $\mathbf q_{{L_2}}(\mathbf X_{\mathrm{nom}})$ and  $\mathbf q_{\mathrm{bf}}(\mathbf X_{\mathrm{nom}})$. For reference, we also plot alongside the true maximum-variance line.
All three lines   visually coincide, i.e.  $\mathbf q_{L_2}(\mathbf X_{\mathrm{nom}})$ and  $\mathbf q_{\mathrm{bf}}(\mathbf X_{\mathrm{nom}})$ are excellent estimates of the true maximum-variance line.

Next, we assume   that instead of the clean matrix of nominal training data points $\mathbf X_{\mathrm{nom}}$, we are given the outlier-corrupted data matrix $\mathbf X_{\mathrm{cor}} = [\mathbf X_{\mathrm{nom}}, \mathbf O] \in \mathbb R^{2 \times 104}$,  for estimating $\mathbf q_{\mathrm{opt}}$. Here, in addition to the previous $100$ nominal data points in $\mathbf X_{\mathrm{nom}}$, $\mathbf X_{\mathrm{cor}}$ also contains the $4$  outliers  in $\mathbf O = [\mathbf o_1, \mathbf o_2, \mathbf o_3, \mathbf o_4] \in \mathbb R^{2 \times 4}$ as seen in Fig. \ref{ch1_fig:linefit2} that do  not follow the nominal  distribution. 
 Similar to our treatment of $\mathbf X_{\mathrm{nom}}$, we calculate  the L2-PC  and L1-PC (by L1-BF) of $\mathbf X_{\mathrm{cor}}$, $\mathbf q_{L_2}(\mathbf X_{\mathrm{cor}})$ and  $\mathbf q_{\mathrm{bf}}(\mathbf X_{\mathrm{cor}})$,  respectively.
 In Fig. \ref{ch1_fig:linefit2} we plot the two lines, against the true maximum-variance line  of the nominal data. 
  This time,  the lines of $\mathbf q_{L_2}(\mathbf X_{\mathrm{cor}})$ and  $\mathbf q_{\mathrm{bf}}(\mathbf X_{\mathrm{cor}})$ differ significantly. In sharp contrast to  L1-PC,  the L2-PC of $\mathbf X_{\mathrm{cor}}$ is strongly attracted by the four outliers  and drifting  away from the true maximum-variance line. 
 The outlier-resistance and superior line-fitting performance of the proposed  approximate   L1-PC in the presence of the faulty data is  clearly illustrated.

\subsection{Classification of Genomic Data -- Prostate Cancer Diagnosis}

In this experiment, we conduct subspace-classification of MicroRNA (miRNA) data for prostate cancer diagnosis \cite{lu, gaur}. Specifically, we operate on the expressions of $D=9$ miRNAs (miR-26a, miR-195, miR-342-3p, miR-126*, miR-425*, miR-34a*, miR-29a*, miR-622, miR-30d) that have been recently considered to be differentially expressed between malignant and  normal human prostate tissues \cite{carlsson}. For each of the above miRNAs, we obtain one sample expression from one malignant and one normal prostate tissue from $N=19$ patients (patient indexes per \cite{carlsson}: $1,2,\ldots,10,12,\ldots, 20$) of the of the Swedish Watchful Waiting cohort \cite{johansson}.\footnote{More information about the studied data can be found in \cite{carlsson}.} The miRNA expressions from malignant and normal tissues are organized in $\mathbf X_{M}=[\mathbf x_{M,1}, \mathbf x_{M,2}, \ldots, \mathbf x_{M,N}] \in \mathbb R^{D \times N}$ and  $\mathbf X_{N} =[\mathbf x_{N,1}, \mathbf x_{N,2}, \ldots, \mathbf x_{N,N}] \in \mathbb R^{D \times N}$, respectively. The classification/diagnosis experiment is conducted as follows. 

\begin{figure}[h]
	\centering
	\includegraphics[width=0.8\linewidth]{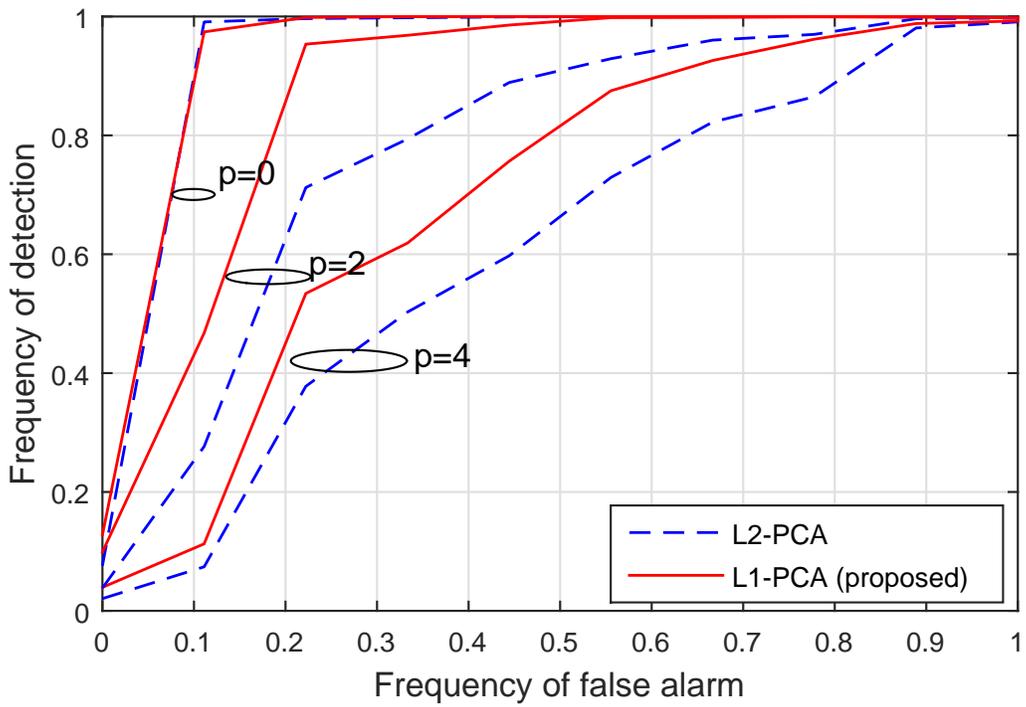}
	\caption{ROC curve of L2-PCA and L1-PCA (L1-BF) malignant tissue detector for $p=0,2,4$ mislabeled points per training dataset ($N=19$, $N_{train}=10$, $D=9$, $K=3$).  \vspace{-0.5cm}}
	\label{diagnosis}
	
\end{figure}

We collect $N_{train}=10$ points from $\mathbf X_{M}$ in $\mathbf X_{M,tr} = [\mathbf X_{M}]_{:,\mathcal I_{M}}$  with $\mathcal I_M \subset \{1, 2, \ldots, N\}$ and $|\mathcal I_M|=N_{train}$, and $N_{train}$ points from $\mathbf X_{N}$ in $\mathbf X_{N,tr} = [\mathbf X_{N}]_{:,\mathcal I_{N}}$ with $\mathcal I_N \subset \{1, 2, \ldots, N\}$ and $|\mathcal I_N|=N_{train}$. 
Then, we calculate the zero-centered malignant and normal tissue training datasets
$\tilde{\mathbf X}_{M,tr} = {\mathbf X}_{M,tr} - \mathbf m_M \mathbf 1_{N_{train}}^\top$ where $\mathbf m_{M} = \frac{1}{N_{train}} {\mathbf X}_{M,tr} \mathbf 1_{N_{train}}$  and 
$\tilde{\mathbf X}_{N,tr} = {\mathbf X}_{N,tr} - \mathbf m_N \mathbf 1_{N_{train}}^\top$ where  $\mathbf m_{N} = \frac{1}{N_{train}} {\mathbf X}_{N,tr} \mathbf 1_{N_{train}}$, respectively. Next, we find  $K=3$ L2-PCs or L1-PCs by L1-BF of the zero-centered malignant and normal datasets, $\mathbf Q_{M} \in \mathbb R^{D \times K}$ and $\mathbf Q_{N} \in \mathbb R^{D \times K}$, respectively, and  conduct subspace-classification  \cite{jain,liu} of each  data-point $\mathbf x$ that has not been used for PC training (i.e., every  $\mathbf x \in \{ \mathbf x_{M,i} \}_{i \in \mathcal I_M^c} \cup \{ \mathbf x_{N,i} \}_{i \in \mathcal I_N^c}$ where $\mathcal I_{M}^c = \{1, 2, \ldots, N \}\setminus \mathcal I_M$ and $\mathcal I_{N}^c = \{1, 2, \ldots, N \}\setminus \mathcal I_N$)  as 
\begin{align}
\| \mathbf Q_{M}^\top (\mathbf x -\mathbf m_M)  \|_2^2 \overset{\overset{\text{\footnotesize{malignant}}}{\text{\normalsize{$>$}}}}{\underset{\text{\footnotesize{normal}}}{{<}}} \| \mathbf Q_{N}^\top (\mathbf x -\mathbf m_N)  \|_2^2 + \lambda 
\end{align}
where $\lambda$ is the classification/detection bias term. 
We run the above classification  experiment over $2 000$ distinct training-dataset configurations 
 and calculate the receiver operating characteristic (ROC) curve of the developed L2-PCA or L1-PCA classifier identifying as ``detection" the event of classifying correctly a malignant tissue and as ``false alarm" the event of classifying erroneously a normal tissue (Fig. \ref{diagnosis}, ``$p=0$" curves).

Next, we consider the event of training on faulty/mislabeled data. That is, we recalculate the above ROC curves having  exchanged $p=2$ or $4$ data points between $\mathbf X_{M,tr}$ and $\mathbf X_{N,tr}$. 
In Fig. \ref{diagnosis}, we plot the ROC curves for the designed L2-PCA and L1-PCA malignant tissue detectors, for $K=3$ and $p=0,2,4$. We observe that for $p=0$ (no mislabeling) the the two classifiers perform almost identically, attaining frequency of detection (FD) close to 1 for frequency of false alarm (FFA) less than 0.15. However, in the event of training data mislabeling, the strong resistance of the proposed L1-PCs against dataset contamination is apparent.
The L1-PCA classifier outperforms significantly  L2-PCA, for every examined value of $p$. For instance, for $p=2$, L1-PCA achieves FD of $.95$ for FFA $0.23$, while L2-PCA achieves the same FD for FFA $.65$.

\section{Conclusions}

\label{ch2_sec7}

In this work, we presented   a novel, near-optimal, bit-flipping based  algorithm named L1-BF,  
to calculate  $K$ L1-PCs of any rank-$d$ $D\times N$ data matrix with 
 complexity $\mathcal O (ND \mathrm{min} \{ N,D\} + N^2(K^4 + dK^2) + NdK^3 )$.
 Formal proof of convergence, theoretical analysis of converging points, and detailed asymptotic complexity derivation were carried out. 
 Our numerical experiments show that the proposed algorithm
 outperforms in the optimization metric all other L1-PCA calculators at cost comparable 
 to L2-PCA.
 L1-PCA and L2-PCA have almost indistinguishable behavior on clean, nominal data. L1-PCA shows remarkable relative resistance to faulty data contamination. 
Thus, the proposed algorithm,  retaining the robustness of L1-PCA at a cost close to that of L2-PCA, may bridge the gap between outlier resistant and computationally efficient principal-component analysis.

\appendix

\subsection{Proof of Proposition \ref{ch2_bounds}}
	Since $\mathbf b_{\mathrm{bf}} \in \Omega(\mathbf Y)$, \eqref{ch2_sbfset2}  and \eqref{ch2_XtoY} imply
	\begin{align}
		\left\|  \mathbf Y \mathbf b_{\mathrm{bf}} \right\|_{2}^2 &= \sum_{n\in \{1, 2, \ldots, N \}} b_n\mathbf y_n^\top\mathbf Y \mathbf b  \nonumber \\
		& \geq  \sum_{n\in \{1, 2, \ldots, N \}}  \left\| \mathbf y_{n}\right\|_2^2 =\left\| \mathbf Y\right\|_F^2 = \left\| \mathbf X\right\|_F^2.
		\label{ch2_SBFbound1}
	\end{align}
	In addition, by Proposition \ref{ch2_relationship},  $\mathbf b_{\mathrm{bf}}$ lies in $\Phi (\mathbf Y)$ as well and satisfies $\mathbf b_{\mathrm{bf}} = \mathrm{sgn}(\mathbf Y^\top \mathbf Y \mathbf b_{\mathrm{bf}})$. Therefore,  for    $\mathbf q_{\mathrm{bf}}  = \mathbf X \mathbf b_{\mathrm{bf}}\left\| \mathbf X \mathbf b_{\mathrm{bf}} \right\|_2^{-1}$, 
	\begin{align}
		\left\| \mathbf X^\top \mathbf q_{\mathrm{bf}}\right\|_1 &=  {\left\|\mathbf X^\top \mathbf X \mathbf b_{\mathrm{bf}}\right\|_1}{\left\| \mathbf X \mathbf b_{\mathrm{bf}}\right\|_2^{-1}} \nonumber \\
		& =  \left( \mathrm{sgn}(\mathbf X^\top \mathbf X \mathbf b_{\mathrm{bf}})^\top \mathbf X^\top \mathbf X \mathbf b_{\mathrm{bf}}  \right)\left\| \mathbf X \mathbf b_{\mathrm{bf}}\right\|_2^{-1} \nonumber \\
		&= \left( \mathrm{sgn}(\mathbf Y^\top \mathbf Y \mathbf b_{\mathrm{bf}})^\top \mathbf X^\top \mathbf X \mathbf b_{\mathrm{bf}}  \right)\left\| \mathbf X \mathbf b_{\mathrm{bf}}\right\|_2^{-1} \nonumber \\
		& = \left\| \mathbf X \mathbf b_{\mathrm{bf}}\right\|_2 \overset{\eqref{ch2_SBFbound1}}{=} \left\| \mathbf Y \mathbf b_{\mathrm{bf}}\right\|_2   \geq \left\| \mathbf X\right\|_F.
		\label{ch2_bound1}
	\end{align}
\hfill $\qed$

\newpage

\bibliographystyle{IEEEbib}

  \newpage

\vspace*{\fill}
\renewcommand{\arraystretch}{1.2}
\begin{figure}[h]
	{\small
		{\hrule height 0.2mm} 
		\vspace{0.3mm}
		{\hrule height 0.2mm}
		\vspace{1mm}
		{\bf L1-BF for the calculation of the first L1-norm principal component ($K=1$)}
		\vspace{0.5mm}
		{\hrule height 0.2mm}
		\vspace{2mm}
		\textbf{Input:}  $\mathbf X_{D \times N}$   \\
		\begin{tabular}{r l }
			1: & $(\mathbf U, \mathbf \Sigma, \mathbf V) \leftarrow \mathrm{csvd} (\mathbf X)$, $\mathbf Y \leftarrow \mathbf \Sigma \mathbf V^\top$  \\
			2: & $ \mathbf b_{\mathrm{bf}}  \leftarrow \mathrm{bf} \left(  \mathbf Y^\top \mathbf Y, \mathrm{sgn}([\mathbf Y]_{:,1}) \right)$ \\
			3: & $\mathbf q_{\mathrm{bf}} \leftarrow \mathbf X \mathbf b_{\mathrm{bf}} /\left\| \mathbf X \mathbf b_{\mathrm{bf}} \right\|_2$
		\end{tabular} \\
		\textbf{Output:}  $\mathbf q_{\mathrm{bf}}$
		\vspace{1.5mm}
		{\hrule height 0.2mm} 
		\vspace{1.5mm}
		{Function } $\mathrm{bf} (\mathbf A, \mathbf b  )$
		\vspace{0.5mm}
		{\hrule height 0.2mm}
		\vspace{1mm}
		\begin{tabular}{r l }
			1: &$N \leftarrow \mathrm{length(\mathbf b)}$, $\mathcal L \leftarrow \{1,2, \ldots, N\}$\\
			2: &  $\alpha_i  \leftarrow  2 b_{i} \sum_{m \neq i}  \,  b_m [\mathbf A]_{i,m} 
			~~\forall i \in \{1, 2, \ldots, N\}$ \\ 
			3: & $~~~$  \text{while   true (or terminate at $N$ BFs)}\\
			4: & $~~~$ $~~$ $n \leftarrow  {\mathrm{argmin}}_{m \in \mathcal L} \alpha_m $\\
			5: & $~~~$ $~~$  \text{if} $a_n \leq 0$, \\
			6: & $~~~$ $~~$ $~~$  $b_n \leftarrow - b_{n}$, $\mathcal L \leftarrow \mathcal L \setminus \{n\}$  \\
			7: &  $~~~$ $~~$ $~~$ $a_n \leftarrow -a_n$, $\alpha_i \leftarrow \alpha_i - 4 b_i b_n [\mathbf A]_{i,n} ~~\forall i\neq n$ \\
			8: & $~~~$ $~~$  \text{elseif}  $a_n > 0$  \text{and} $|\mathcal L| < N$, $\mathcal L \leftarrow \{1, 2, \ldots, N \}$ \\
			9: & $~~~$ $~~$  \text{else}, \text{break} \\
			10: &    \text{Return}  ${\mathbf b} $
		\end{tabular} 
		\vspace{1mm}
		{\hrule height 0.2mm} 
		\vspace{0.3mm}
		{\hrule height 0.2mm}
		\vspace{0.6cm}
	}
	\caption{The proposed L1-BF  algorithm for the calculation of the    L1-principal component of a rank-$d$ data matrix $\mathbf X_{D \times N}$ of $N$ samples of dimension $D$; $\mathrm{csvd}(\cdot)$ returns the compact SVD of the argument. }
	\vspace{-0.4cm}
	\label{ch2_algo1}
\end{figure}%

\vspace*{\fill}
\newpage
\vspace*{\fill}

\vspace*{\fill}
\newpage
\vspace*{\fill}

\vspace*{\fill}
\newpage
\vspace*{\fill}

\vspace*{\fill}
\newpage
\vspace*{\fill}

\renewcommand{\arraystretch}{1.5}

\begin{table}[h]
	\caption{Computational cost of L1-BF for the calculation of the principal component ($K=1$) of a $D \times N$ real data matrix.}
	\vspace{0.3cm}
	\centering
	\begin{tabular}{ | l| l |}
		\hline 
		Computational Task & Computational Cost \\
		\hline  
		$\mathbf Y$ and $\mathbf b^{(1)}$ & $\mathcal O(ND ~\mathrm{min}\{N,D \})$ \\ 
		$\mathbf Y^\top \mathbf Y$ & $\mathcal O(N^2d)$ \\
		One bit flip  & $\mathcal O(N)$ \\
		$N$ BF iterations & $\mathcal O(N^2)$  \\
		$\mathbf q_{\mathrm{bf}}$ from $\mathbf b_{\mathrm{bf}}$ & $\mathcal O(ND)$   \\
		\hline 
		Total & $\mathcal O(ND~\mathrm{min}\{N,D\}  + N^2d)$  \\
		\hline 
	\end{tabular}
	\label{ch2_table:complexity}
\end{table}
\renewcommand{\arraystretch}{1}

\vspace*{\fill}
\newpage
\vspace*{\fill}

%
%
%
%
%

\renewcommand{\arraystretch}{1.2}

\begin{figure}[h]
{\small
{\hrule height 0.2mm} 
\vspace{0.3mm}
{\hrule height 0.2mm}
\vspace{1mm}
{\bf  L1-BF for the calculation of $K > 1$ L1-norm principal components}
\vspace{0.5mm}
{\hrule height 0.2mm}
\vspace{2mm}
\textbf{Input:} Data matrix $\mathbf X_{D \times N}$ of rank $d$, $K  \leq d$  \\
\begin{tabular}{r l }
1: & $\left( \mathbf U , \mathbf \Sigma_{d \times d}, \mathbf V  \right) \leftarrow \mathrm{csvd}(\mathbf X)$\\
2: & $\mathbf Y \leftarrow \mathbf \Sigma \mathbf V^\top$,  $\mathbf v \leftarrow [\mathbf V]_{:,1}$, $\mathbf B = \mathrm{sgn}(\mathbf v \mathbf 1_{K}^\top)$\\
3: & $\mathbf B_{\mathrm{bf}} \leftarrow \mathrm{bfK} \left( \mathbf Y, \mathbf B , K,   \right)$ \\
4: & $\left( \hat{\mathbf U}_{D \times K}, \hat{\mathbf \Sigma}_{K \times K}, \hat{\mathbf V}_{K \times K} \right) \leftarrow \mathrm{svd}(\mathbf X \mathbf B_{\mathrm{bf}})$\\
5: &$\mathbf Q_{\mathrm{bf}} \leftarrow \hat{\mathbf U}\hat{\mathbf V}^\top$  
\end{tabular} \\
\textbf{Output:}  $\mathbf Q_{\mathrm{bf}}  $  
\vspace{1.5mm}
{\hrule height 0.2mm}
\vspace{1.5mm}
 Function  $\mathrm{bfK} (\mathbf Y_{d \times N}, \mathbf B_{N \times K}, K \leq d) $
\vspace{0.5mm}
{\hrule height 0.2mm}
\vspace{1mm}
\begin{tabular}{r l }
1: &  $\omega \leftarrow K \| \mathbf Y [\mathbf B]_{:,1}\|_2$ \\
2: &  $\mathcal L \leftarrow \{1,2 , \ldots, K \}$  \\
3: &  \text{while true (or terminate at $NK$ BFs)}  \\
4: &  $~~~$   $(\mathbf U, \mathbf S_{K \times K}, \mathbf V^\top) \leftarrow \mathrm{svd} (\mathbf Y \mathbf B)$, $\mathbf F \leftarrow \mathbf U \mathbf S$ \\
5: &  $~~~$   \text{for} $x \in \mathcal L$, $m \leftarrow \mathrm{mod}(x, N)$, $ l \leftarrow (x-m)/N + 1$\\
6: &  $~~~$  $~~~$   $( [\mathbf q_{1}, \mathbf q_{2}],  \mathbf D ) \leftarrow \mathrm{evd} (\| \mathbf y_{m}\|_2^2 \mathbf e_{1,2} \mathbf e_{1,2}^\top + [\mathbf e_{2,2}, \mathbf e_{1,2}])$\\
7: &  $~~~$  $~~~$   $\mathbf W \leftarrow \left[ [\mathbf V]_{l,:}^\top, -2 B_{m,l} \mathbf F^\top \mathbf y_{m} \right]$   \\
8: &  $~~~$  $~~~$   $(\mathbf Z, \mathrm{diag}(\mathbf p)) \leftarrow \mathrm{fevd} (\mathbf S^\top \mathbf S + d_{1} \mathbf W \mathbf q_{1}(\mathbf W \mathbf q_{1})^\top)$ \\ 
9: &  $~~~$  $~~~$   $(\sim,  \boldsymbol \Lambda ) \leftarrow \mathrm{fevd} (\mathrm{diag}(\mathbf d)+ d_{2} \mathbf Z^\top \mathbf W \mathbf q_{2}(\mathbf Z^\top \mathbf W \mathbf q_{2})^\top)$    \\
10: &  $~~~$  $~~~$   $ a_{l,k} \leftarrow \sum_{j=1:K} \sqrt{\lambda_{j}}$  \\ 
11: &  $~~~$    $(n,k) \leftarrow \mathrm{argmax}_{m,l:~(l-1)N+m \in \mathcal L} ~a_{m,l}$  \\ 
12: &  $~~~$    \text{if } $\omega < a_{n,k}$, \\
13: &  $~~~$  $~~~$ $B_{n,k} \leftarrow - B_{n,k}$, $\omega \leftarrow a_{n,k}$,\\
14: &  $~~~$  $~~~$ $\mathcal L \leftarrow \mathcal L \setminus \{ (k-1)N+n \}$ \\ 
15: &  $~~~$    \text{elseif }  $\omega \geq a_{n,k}$ \text{and} $|\mathcal L| < NK$, $\mathcal L \leftarrow \{1, 2, \ldots, NK \}$ \\ 
16: &  $~~~$    \text{else, break } \\ 
17: &   \text{Return }  $\mathbf B$ \\ 
\end{tabular} 
\vspace{0.6mm}
{\hrule height 0.2mm} 
\vspace{0.3mm}
{\hrule height 0.2mm}
\vspace{0.4cm}
} 
\caption{The proposed  L1-BF   algorithm for the calculation of    $K >1$ L1-principal components of a rank-$d$ data matrix $\mathbf X_{D \times N}$ of $N$ samples of dimension $D$;  $\mathrm{csvd}(\cdot)$ returns the compact SVD of the argument;  $\mathrm{fevd}(\cdot)$ returns the  EVD of the argument (diagonal positive semidefinite matrix perturbated by a rank-$1$ symmetric) by the algorithm of \cite{GolubSVD}.}
\vspace{-0.4cm}
\label{ch2_algo2}
\end{figure}%

\renewcommand{\arraystretch}{1}

\vspace*{\fill}
\newpage
\vspace*{\fill}

\renewcommand{\arraystretch}{1.5}
\begin{table}[h]
	\caption{Computational complexity of L1-BF for the calculation of  $K>1$ principal components of a $D \times N$ real data matrix of rank $d$.}
	\vspace{0.3cm}
	\centering
	\begin{tabular}{ | l| l |}
		\hline 
		Computational Task & Computational Cost \\
		\hline  
		$\mathbf Y$ and $  \mathbf B^{(1)}$ & $\mathcal O(ND \mathrm{min}\{N,D \})$ \\ 
		One bit flip & $\mathcal O(dK^2 + N(K^3 + dK) )$ \\ 
		$NK$  BF iterations & $ \mathcal O (N^2(K^4 + dK^2) NdK^3 )$ \\
		$\mathbf Q_{\mathrm{bf}}$ by $\mathbf B_{\mathrm{bf}}$ & $\mathcal O(ND\mathrm{min}\{N,D \} + NDK)$   \\
		\hline 
		Total & $\mathcal O (ND \mathrm{min} \{ N,D\}  + N^2(K^4 + dK^2) + NdK^3)$\\
		\hline 
	\end{tabular}
	\vspace{0.4cm}
	\label{ch2_table:complexity2}
\end{table}
\renewcommand{\arraystretch}{1}

\vspace*{\fill}

\newpage 

\vspace*{\fill}

\vspace*{\fill}
\newpage
\vspace*{\fill}

\vspace*{\fill}
\newpage
\vspace*{\fill}

\vspace*{\fill}
\newpage
\vspace*{\fill}

\vspace*{\fill}

\end{document}